\documentclass{ws-rv961x669}
\usepackage{graphicx}
\usepackage[utf8]{inputenc}
\usepackage{hyperref}
\usepackage{color}
\usepackage{slashed}
\usepackage{hyperref}
\usepackage[square]{ws-rv-van}
\newcommand{\beq}{\begin{equation*}}
\newcommand{\eeq}{\end{equation*}}
\newcommand{\betabeta}{0\nu\beta\beta}
\newcommand{\hbeta}{$\mbox{}^3 {\rm H}$ $\beta$-decay \ }

\def\ltap{\ \raisebox{-.4ex}{\rlap{$\sim$}} \raisebox{.4ex}{$<$}\ }
\def\gtap{\ \raisebox{-.4ex}{\rlap{$\sim$}} \raisebox{.4ex}{$>$}\ }
\newcommand{\eV}{\mbox{$ \  \mathrm{eV}$}}

\def\cropnote{\relax}

\begin{document}

\chapter*{CP Violation in the Lepton Sector and  Implications for Leptogenesis \label{CH6}}


\author[]{C.~Hagedorn$^*$, R.~N.~Mohapatra$^\dagger$, E.~Molinaro$^*$\footnote{Corresponding Author.}, C.~C.~Nishi$^\ddagger$, S.~T.~Petcov$^\S$\footnote{Also at Institute of Nuclear Research and Nuclear Energy, Bulgarian Academy of Sciences, 1784 Sofia, Bulgaria.}}

\address{$^*$CP$^3$-Origins, University of Southern Denmark,
Campusvej 55,\\ DK-5230 Odense M, Denmark
\\[3pt]
$^\dagger$Maryland Center for Fundamental Physics, Department of Physics,\\ University of Maryland,
College Park, MD 20742, USA
\\[3pt]
$^\ddagger$Universidade Federal do ABC, Centro de Matem\'atica,
Computa\c{c}\~{a}o e Cogni\c{c}\~ao Naturais, 09210-580, Santo Andr\'e-SP, Brasil
\\[3pt]
$^\S$SISSA/INFN, Via Bonomea 265, 34136 Trieste, Italy, 
\\[3pt]
Kavli IPMU (WPI), University of Tokyo, 5-1-5 Kashiwanoha,\\ 
277-8583 Kashiwa, Japan
}

\begin{abstract}
\textbf{Abstract}: We review the current status of the 
data on neutrino masses and lepton mixing 
and the prospects for measuring the CP-violating  phases 
in the lepton sector. The possible connection between low energy 
CP violation encoded in the Dirac and Majorana phases of 
the Pontecorvo-Maki-Nakagawa-Sakata mixing matrix 
and successful  leptogenesis is emphasized in the context of
seesaw extensions of the Standard Model with a 
flavor symmetry $G_f$ (and CP symmetry).
\end{abstract}


\body

\tableofcontents

\section{Introduction: the three-neutrino mixing scheme}
\label{aba:sec2}
%

The last 19 years witnessed remarkable discoveries in the field 
of neutrino physics, which point towards new and yet unknown physics beyond the Standard Model (SM). The  solar and atmospheric neutrino experiments 
and the experiments with reactor and accelerator neutrinos
have provided irrefutable proofs of  
neutrino oscillations \cite{Pontecorvo:1957cp,Maki:1962mu,Pontecorvo:1967fh} --
transitions in flight --
between the different flavor neutrinos
$\nu_e$, $\nu_\mu$, $\nu_\tau$
(antineutrinos $\bar{\nu}_e$, $\bar{\nu}_\mu$, $\bar{\nu}_\tau$) 
caused by nonzero neutrino masses and lepton mixing 
(see, e.g.,  \cite{PDG2016} for a review of the relevant data).
The experimental discovery of oscillations of atmospheric 
muon neutrinos and antineutrinos and of the flavor conversion of 
solar (electron) neutrinos led to the 2015 Nobel Prize for Physics 
awarded to Takaaki Kajita \cite{NPP15TKajita} (from the SuperKamiokande Collaboration) 
and Arthur B. McDonald \cite{BPP15AMcDonald} (from the SNO Collaboration). 
The existence of flavor neutrino oscillations implies 
the presence of mixing in the lepton weak charged current (CC), namely
\begin{equation}
\label{CC}
{\cal L}_{\rm CC} = - ~\frac{g}{\sqrt{2}}\,
\sum_{\alpha=e,\mu,\tau}
\bar{e}_{L\alpha}\, \gamma_{\mu} \nu_{L \alpha}\,
W^{\mu \dagger} + \text{h.c.}\,,\quad\quad
\nu_{L\alpha}
= \sum^n_{j=1} \left(U_{\rm PMNS}\right)_{\alpha j} \,\nu_{j L}\,,
\end{equation}
\noindent 
where $\nu_{L \alpha}$ is the flavor neutrino field, 
$\nu_{j L}$ is the left-handed (LH)
component of the field of the neutrino $\nu_j$ having a 
mass $m_j$, and $U_{\rm PMNS}$ is a unitary matrix -- the
Pontecorvo-Maki-Nakagawa-Sakata (PMNS)
mixing matrix \cite{Pontecorvo:1957cp,Maki:1962mu,Pontecorvo:1967fh}, 
$U_{\rm PMNS}\equiv U_\nu$.\footnote{We work here in a basis where all the lepton mixing 
originates from the neutrino sector, unless stated otherwise.
}

All well established statistically significant data on neutrino 
oscillations can be described within the three-neutrino 
mixing scheme, $n=3$.
The number of neutrinos with definite mass 
can, in general, be bigger than three if, e.g., 
there exist right-handed (RH)  neutrinos \cite{Pontecorvo:1967fh}
and they mix effectively 
(via Majorana and Dirac mass terms)
with the LH flavor neutrinos. 
It follows from the current data that at least three of 
the neutrinos with definite mass $\nu_j$, say 
$\nu_1$, $\nu_2$, $\nu_3$, must be light, with
$m_{1,2,3} \ltap 1$ eV, and must have different 
masses, $m_1\neq m_2 \neq m_3$.  
In what concerns the masses of the possible 
additional mass-eigenstate neutrinos, 
$\nu_k$, $k=4,5,\ldots,$ there are 
a number of different possibilities and we refer the reader to the discussions in other chapters of this review \cite{leptogenesis:A02,leptogenesis:A03,leptogenesis:A05}.\footnote{At present there are several 
experimental inconclusive indications 
for the existence of one or two light 
RH neutrinos at the eV scale, 
which mix with the flavor neutrinos, 
implying the presence of   one or two
 additional neutrinos, $\nu_4$ or $\nu_{4,5}$, 
with masses $m_4~(m_{4,5})\sim 1~{\rm eV}$     
(see, e.g.,  \cite{PDG2016}).
For a discussion of the current status of these 
indications, of their upcoming tests and of the 
related implications see, e.g., \cite{Kopp:2013vaa,Gariazzo:2015rra,Collin:2016rao,Capozzi:2016vac,Gariazzo:2017fdh}.
}

 In the case of three light neutrinos, $n=3$, 
the three-by-three unitary lepton mixing 
matrix $U_\nu$ can be parametrized, as is well known, 
by three angles and one Dirac phase for Dirac neutrinos, and with two additional Majorana phases in case neutrinos are Majorana particles \cite{Bilenky:1980cx}, that is
\begin{equation}
U_\nu= \hat{U}_\nu \cdot {\rm diag}(1, e^{i \alpha}, e^{i \beta})\,, 
\label{VP}
\end{equation}
%
where $\alpha$  and $\beta$
are the two Majorana phases, $\alpha,\beta\in[0,\pi)$, and 
\begin{equation} 
\begin{array}{c}
\label{eq:Vpara}
\hat{U}_\nu = \left(\begin{array}{ccc} 
 c_{12} c_{13} & s_{12} c_{13} & s_{13} e^{-i \delta}  \\[0.2cm] 
 -s_{12} c_{23} - c_{12} s_{23} s_{13} e^{i \delta} 
 & c_{12} c_{23} - s_{12} s_{23} s_{13} e^{i \delta} 
 & s_{23} c_{13} 
\\[0.2cm] 
 s_{12} s_{23} - c_{12} c_{23} s_{13} e^{i \delta} & 
 - c_{12} s_{23} - s_{12} c_{23} s_{13} e^{i \delta} 
 & c_{23} c_{13} 
\\ 
  \end{array}    
\right)\,. 
\end{array} 
\end{equation}
%
\noindent 
In \eref{eq:Vpara},
$c_{ij} = \cos\theta_{ij}$, 
$s_{ij} = \sin\theta_{ij}$,
the angles $\theta_{ij} \in [0,\pi/2)$, 
and $\delta \in [0,2\pi)$ is the 
Dirac  phase. 
Thus, if $\nu_j$ are Dirac particles, the PMNS mixing matrix 
is similar, in both the number of 
mixing angles and CP phases, to the 
quark mixing matrix. 
 $U_\nu$ contains two additional 
physical CP phases, if  $\nu_j$ are Majorana 
fermions due to the special properties of Majorana 
particles (see, e.g., \cite{Bilenky:1980cx,Bilenky:1987ty,Petcov:2013poa}).
On the basis of the currently existing neutrino data 
it is impossible to determine 
the nature -- Dirac or Majorana -- of the neutrinos 
with definite mass~$\nu_j$.\\

 The  neutrino oscillation  probabilities
are functions of the neutrino energy, $E$, 
of the source-detector distance $L$, 
of the elements of $U_\nu$ and, 
for relativistic neutrinos used in all neutrino 
experiments performed so far, 
of the neutrino mass squared 
differences $\Delta m^2_{ij} \equiv m^2_{i} - m^2_j$, $i\neq j$ 
(see, e.g., \cite{PDG2016,Bilenky:1987ty}).
In the three-neutrino mixing scheme,
there are only two independent $\Delta m^2_{ij}$,
say $\Delta m^2_{21}\neq 0$ and $\Delta m^2_{31} \neq 0$.
In the widely used convention of numbering 
the neutrinos with definite mass $\nu_j$ that
we are going to employ, 
$\theta_{12}$, $\Delta m^2_{21}> 0$, and
$\theta_{23}$, $\Delta m^2_{31}$,  
represent the parameters which drive the 
solar ($\nu_e$) and the dominant atmospheric 
($\nu_{\mu}$ and $\bar{\nu}_{\mu}$) 
oscillations, respectively, 
and  $\theta_{13}$ is associated
with the smallest mixing angle in 
the PMNS mixing matrix.
The existing  data allows a determination of 
$\theta_{12}$, $\Delta m^2_{21}$,  and
 $\theta_{23}$, $|\Delta m^2_{31(32)}|$ and $\theta_{13}$, 
with a few percent 
precision \cite{Capozzi:2017ipn,Esteban:2016qun,deSalas:2017kay}.
The best fit value (bfv) and the $3\, \sigma$ allowed ranges of 
$\Delta m^2_{21}$, $\sin^2\theta_{12}$, $|\Delta m^2_{31(32)}|$, 
$\sin^2\theta_{23}$ and $\sin^2\theta_{13}$ read \cite{Capozzi:2017ipn}
\begin{eqnarray}
\begin{split}
\label{deltasolvalues}
&(\Delta m^2_{21})_{\rm bfv} = 7.37 \times 10^{-5}\ \eV^2,~~~~~~ 
 \Delta m^2_{21} = (6.93 \div 7.96) \times 10^{-5} \ \eV^2\,,\\
\label{sinsolvalues}
& (\sin^2 \theta_{12})_{\rm bfv} = 0.297,~~~~~~
 0.250 \leq \sin^2 \theta_{12} \leq 0.354\,,
~~~~~~~~~~~~~~~~~~~~\\ 
\label{deltaatmvalues}
&(|\Delta m^2_{31(32)}|)_{\rm bfv} = 2.56~(2.54) \times 10^{-3} \ \eV^2\,,
~~~~~~~~~~~~~~~~~~~~~~~~~~~~~~~~~~~~~~~\\ 
& |\Delta m^2_{31(32)}| = (2.45~(2.42) \div 2.69~(2.66))\times 10^{-3}\ \eV^2\,,
~~~~~~~~~~~~~~~~~~~~~~~~~~~~~~~~~\\ 
 \label{thetaatmvalues}
&(\sin^2\theta_{23})_{\rm bfv} = 0.425~(0.589)\,,~~ 
 0.381\,(0.384) \leq \sin^2\theta_{23} \leq 0.615\,(0.636)\,,
~~~~~~~~~\\
 \label{theta13values}
&(\sin^2\theta_{13})_{\rm bfv} = 0.0215~(0.0216)\,,~~ 
0.0190\,(0.0190) \leq \sin^2\theta_{13} \leq 0.0240\,(0.0242)\,,
\end{split}
\end{eqnarray}
%
where
the value (the value in brackets) 
corresponds to  $\Delta m^2_{31(32)}>0$ ($\Delta m^2_{31(32)} <0$).
The current data does not fix the sign of  $\Delta m^2_{31(32)}$.
It follows from the results quoted above that 
$\Delta m^2_{21}/|\Delta m^2_{31(32)}| \approx 0.03$ and
$|\Delta m^2_{31}| = |\Delta m^2_{32} + \Delta m^2_{21}| 
\approx |\Delta m^2_{32}|$.  Notice also that the atmospheric mixing 
angle can be maximal, $\theta_{23}=\pi/4$, within the $3\, \sigma$
interval. 
In contrast, the value of $\theta_{12} = \pi/4$, i.e.,
maximal solar neutrino mixing,  
is definitely ruled out 
by the data. Thus, one has  $\theta_{12} < \pi/4$ and 
at $99.73\%$ CL, $\cos2\theta_{12} \geq 0.29$.
The quoted results imply also that 
the pattern of lepton mixing differs 
drastically from the pattern of quark mixing. 

Starting from 2013 there are also persistent 
hints in the neutrino oscillation data that the value of the Dirac phase
$\delta\approx 3\pi/2$ (see, e.g.,  
\cite{Capozzi:2013csa,Gonzalez-Garcia:2014bfa}).\footnote{Using the 
T2K data on $\nu_{\mu}\rightarrow \nu_e$ oscillations, 
the  T2K Collaboration \cite{Abe:2013hdq,T2K:Nu2016}
found  in 2013  that
for $\delta = 0$, $\sin^2\theta_{23} = 0.5$ and 
$|\Delta m^2_{31(32)}| = 2.4\times 10^{-3}{\rm eV^2}$,
in the case of $\Delta m^2_{31(32)}>0$ 
($\Delta m^2_{31(32)} < 0$), 
$\sin^22\theta_{13} = 0.140^{+0.038}_{-0.032}$ 
($0.170^{+0.045}_{-0.037}$). Thus, the bfv of 
$\sin^22\theta_{13}$  obtained in the T2K experiment 
for $\delta = 0$ was approximately a factor of 1.6 (1.9) 
bigger than that measured in the Daya Bay 
experiment \cite{An:2013zwz}: 
$\sin^22\theta_{13} = 0.090^{+0.008}_{-0.009}$. 
The compatibility of 
the results of the two experiments on $\sin^22\theta_{13}$ 
required, in particular, that  $\delta \neq 0$,
which led to the hint that $\delta\approx 3\pi/2$ 
in the analyses 
\cite{Abe:2013hdq,Capozzi:2013csa,Gonzalez-Garcia:2014bfa} 
of the neutrino oscillation data.
}  
In the two recent analyses 
\cite{Capozzi:2017ipn,Esteban:2016qun} 
the authors find for the bfv of 
$\delta$, respectively,  $\delta=1.38\pi$ ($\delta=1.31\pi$)
and $\delta=1.45\pi$ ($\delta= 1.54\pi$) 
in the case of $\Delta m^2_{31(32)}>0$ ($\Delta m^2_{31(32)} <0$).
According to the results in \cite{Capozzi:2017ipn} 
the CP-conserving value $\delta=0$ 
is disfavored at $2.4\, \sigma$  
(at $3.2\, \sigma$) for  $\Delta m^2_{31(32)}>0$
($\Delta m^2_{31(32)}<0$), while the maximal CP-violating value 
$\delta=\pi/2$ is ruled out at $3.4\, \sigma$ (at $3.9\, \sigma$).
The second CP-conserving value $\delta = \pi$ is statistically 
about $2.0\, \sigma$ away from 
the bfv $\delta=1.38\pi$  
(see, e.g., figure~1 in \cite{Capozzi:2017ipn}).
The hint that $\delta \approx 3\pi/2$ is strengthened 
by the results of the NO$\nu$A neutrino 
oscillation experiment \cite{Adamson:2016tbq, Vahle:2016}.
 The relatively large value of $\sin\theta_{13} \approx 0.15$, 
 measured in the Daya Bay\cite{An:2012eh,An:2013uza,An:2015rpe,Yu:2016}, RENO \cite{Ahn:2012nd,Choi:2015,Joo:2016} 
 and Double Chooz \cite{Abe:2011fz,Abe:2013sxa,Abe:2014bwa} experiments,
 combined with the value of $\delta \approx 3\pi/2$ 
 has far-reaching implications for the searches for 
 CP violation in neutrino oscillations (see discussion in \sref{sec2}).

  As we have noted, 
the sign of $\Delta m^2_{31(32)}$ cannot be determined 
from the existing data. 
 In the three-neutrino mixing scheme,  
the two possible signs of
$\Delta m^2_{31(32)}$ correspond to two 
types of neutrino mass spectrum.
In the convention of numbering 
of neutrinos $\nu_j$ employed by us,
the two spectra read:\\
{\it i) spectrum with normal ordering (NO)}:
$m_1 < m_2 < m_3$, i.e., $\Delta m^2_{31(32)} >0$ and
$\Delta m^2_{21} > 0$;\\~~ 
{\it ii) spectrum with inverted ordering (IO)}:
$m_3 < m_1 < m_2$, i.e., $\Delta m^2_{32(31)}< 0$, 
$\Delta m^2_{21} > 0$,
$m_{2} = \sqrt{m_3^2 + |\Delta m^2_{32}|}$ and
$m_{1} = \sqrt{m_3^2 + |\Delta m^2_{32}| - \Delta m^2_{21}}$.\\ 
Depending on the values of the lightest neutrino mass, 
the neutrino mass spectrum can  be further distinguished:~\\
{\it a) normal hierarchical (NH)}: $m_1 \ll m_2 < m_3$, 
$m_2 \approx \sqrt{\Delta m^2_{21}}\approx 8.6\times 10^{-3}$ eV,
$m_3 \approx \sqrt{\Delta m^2_{31}} \approx 0.051$ eV; \\ 
{\it b) inverted hierarchical (IH)}: $m_3 \ll m_1 < m_2$, $m_{1} \approx \sqrt{|\Delta m^2_{32}| - \Delta m^2_{21}}\approx 0.050$ eV,
$m_{2} \approx \sqrt{|\Delta m^2_{32}|}\approx 0.051$ eV; \\ 
{\it c) quasi-degenerate (QD)}: $m_1 \approx m_2 \approx m_3 \approx m_{\rm QD}$,
$m_j^2 \gg |\Delta m^2_{31(32)}|$, $m_{\rm QD} \gtap 0.10$ eV.\\ 
All three types of spectra are compatible
with the constraints on the absolute scale 
of neutrino masses. Determining the type of neutrino 
mass spectrum is one of the main goals of the future 
experiments in the field of neutrino physics
(see, e.g., \cite{PDG2016,Qian:2015waa}).\footnote{For a review of the experiments which can 
 provide data on the type of neutrino mass spectrum see, e.g.,
 \cite{Qian:2015waa}. 
For some specific proposals see, e.g., \cite{Petcov:2001sy,Pascoli:2002xq,Petcov:2005rv}.
}

Data on the absolute neutrino mass scale 
can be obtained, e.g., 
from measurements of the spectrum 
of electrons near the endpoint in 
\hbeta experiments \cite{Perrin:1933,Fermi:1934sk,Kraus:2004zw,Lobashev:2003kt,Aseev:2011dq}
and from cosmological and astrophysical 
observations \cite{Abazajian:2011dt}. The most stringent upper 
bound on the $\bar{\nu}_e$ mass 
was reported by the Troitzk
experiment~\cite{Aseev:2011dq} 
\beq
m_{\bar{\nu}_e} < 2.05~\mathrm{eV}\quad\mbox{at}\quad 95\%\quad\mathrm{CL}\,. 
\label{H3beta}
\eeq
%
\noindent A similar result 
was obtained in the 
Mainz experiment \cite{Kraus:2004zw}: 
$m_{\bar{\nu}_e} < 2.3~\rm{eV}$ at 95\% CL.
We have $m_{\bar{\nu}_e} \approx m_{1,2,3}$
in the case of QD  spectrum. 
The upcoming KATRIN experiment~\cite{Lobashev:2003kt,Eitel:2005hg}
is designed to reach a sensitivity  
of  $m_{\bar{\nu}_e} = 0.20$~eV,
i.e., to probe the region of the QD spectrum. 

Constraints on the sum of the neutrino masses 
can be obtained from cosmological and astrophysical data 
\cite{Abazajian:2011dt}.
 Depending on the complexity of the underlying model  and the input data used 
one typically obtains
\begin{equation}
\sum_j m_j \; \ltap\;  (0.3 \div 1.3)~\text{eV}\quad\mbox{at}\quad 95\%\quad\mathrm{CL}\,.
\end{equation}
Assuming the existence of
three light massive neutrinos and the validity of 
the $\Lambda$ CDM (Cold Dark Matter) model,
and using  data on the Cosmic Microwave Background (CMB)
temperature power spectrum anisotropies, 
on gravitational lensing effects and the low $\ell$ 
CMB polarization spectrum data (the
“low P” data), the Planck Collaboration\cite{Ade:2015xua} reported the 
updated upper limit: $\sum_j m_j < 0.57$ eV,  at 95\% CL. 
Adding supernovae (SN) light-curve data and data on 
the Baryon Acoustic Oscillations (BAO) strengthens the limit to~\footnote{Result quoted in \href{https://wiki.cosmos.esa.int/planckpla2015/images/0/07/Params\_table\_2015\_limit95.pdf}{wiki.cosmos.esa.int/planckpla2015}, page 311.}
\begin{equation}
\sum_j m_j\, < \, 0.153~\eV\, \quad\mbox{at}\quad 95\%\quad\mathrm{CL}\,.
\label{Planck1}
\end{equation}
%

  Understanding the origin of the observed pattern of lepton mixing, 
establishing the status of the CP symmetry in the lepton sector, 
determining the type of spectrum the neutrino masses obey and 
determining the nature -- Dirac or Majorana -- of massive neutrinos,  
are among the highest priority goals of the program of 
future research in neutrino physics 
(see, e.g., 
\cite{PDG2016,DellOro:2016tmg,Acciarri:2015uup,Acciarri:2016crz,Acciarri:2016ooe,Abe:2015zbg,Abe:2016ero,An:2015jdp,Kim:2014rfa,Aartsen:2014oha,Adrian-Martinez:2016fdl}).
The principal goal is to reveal at a fundamental level
the mechanism giving rise to neutrino masses and lepton mixing.

%
\section{Observables related to low energy CP violation in the lepton sector}
\label{sec2}
%
%

Apart from the hint that 
the Dirac phase $\delta \approx 3\pi/2$,  
no other experimental information on
the CP phases in the lepton
sector is available 
at present. Thus, the status of CP symmetry 
in the lepton sector is essentially unknown. 
The determination of the Dirac and Majorana phases might shed further light 
on the organizing principles in the lepton (and possibly the quark) sector(s) 
(see discussion in \sref{sec4}).

\subsection{Dirac CP violation}

Taking into account that the reactor mixing angle is not so small, $\theta_{13} \approx 0.15$,
a non-zero value of the Dirac phase $\delta$ 
can generate CP-violating effects in neutrino 
oscillations\cite{Cabibbo:1977nk,Bilenky:1980cx,Barger:1980jm},
i.e., a difference between the probabilities of the 
$\nu_\alpha \rightarrow \nu_{\beta}$ and
$\bar{\nu}_\alpha \rightarrow \bar{\nu}_{\beta}$
oscillations, $\alpha\neq \beta$. 
A measure of CP violation is provided, e.g., 
by the asymmetries

\begin{eqnarray}
A^{(\alpha,\beta)}_{\rm CP} = P({\nu_\alpha \rightarrow \nu_{\beta}}) 
- P({\bar{\nu}_\alpha \rightarrow \bar{\nu}_{\beta}})\,,~~
\alpha\neq \beta\quad\text{and}\quad\alpha,\,\beta=e,\mu,\tau\,.
\end{eqnarray}
%
\noindent \\The magnitude of  CP-violating  
effects in neutrino oscillations in the  
three-neutrino mixing scheme is controlled, 
as is well known~\cite{Krastev:1988yu},  
by the rephasing invariant $J_{\rm CP}$  
associated with the Dirac phase $\delta$,

\begin{equation}\label{CPTAsym}
A^{(e,\mu)}_{\rm CP} = A^{(\mu,\tau)}_{\rm CP} = 
- A^{(e,\tau)}_{\rm CP} = J_{\rm CP}~F^{\rm vac}_{\rm osc}\,,
\end{equation}
with
%
\begin{equation}
\label{JCPF}
\begin{split}
J_{\rm CP} & =  {\rm Im}\left( \left(U_\nu\right)_{e2} \left(U_\nu\right)^\ast_{\mu 2} \left(U_\nu\right)_{e 3}^\ast \left(U_\nu\right)_{\mu 3} \right)\,,\\
F^{\rm vac}_{\rm osc} &=
\sin\left(\frac{\Delta m^2_{21}L}{2E}\right) +
\sin\left(\frac{\Delta m^2_{32}L}{2E}\right) +
\sin\left(\frac{\Delta m^2_{13}L}{2E}\right)\,.
\end{split}
\end{equation}

\noindent The  factor $J_{\rm CP}$  in the expressions 
for  $A^{(\alpha,\beta)}_{\rm CP}$ 
is analogous to the rephasing invariant  
associated with the 
Dirac phase in the Cabibbo-Kobayashi-Maskawa (CKM)
quark mixing matrix, introduced in \cite{Jarlskog:1985ht}.
In the standard parametrization of the
PMNS mixing matrix given in \eref{VP} and \eref{eq:Vpara},  $J_{\rm CP}$ has the form

 \begin{equation}
 \begin{split}
 J_{\rm CP}= & \; 
 \frac{1}{8}\,\cos\theta_{13}
 \sin 2\theta_{12}\,\sin 2\theta_{23}\,\sin 2\theta_{13}\,\sin \delta\,.
 \label{JCPstpar}
 \end{split}
 \end{equation}
%
\\ As we discussed, the 
existing neutrino oscillation data has
allowed the determination of the mixing angles 
$\theta_{12}$, $\theta_{23}$ and $\theta_{13}$ 
with a few to several percent  precision.
The size of CP-violating effects in neutrino oscillations 
is still unknown, because the precise
value of the Dirac phase $\delta$ is  currently unknown.
The current data implies 
that in the case of NO (IO) neutrino mass spectrum

\begin{equation}
 0.026~(0.027)\,|\sin\delta| \ltap |J_{\rm CP}| \ltap  0.036\, |\sin\delta|\,,
\label{JCP3sigma}
\end{equation}

\noindent \\ where we have used the $3\, \sigma$ ranges of 
$\sin^2\theta_{12}$, $\sin^2\theta_{23}$ and $\sin^2\theta_{13}$
given in \eref{sinsolvalues}. 
For the  bfv
of $\sin^2\theta_{12}$, $\sin^2\theta_{23}$, $\sin^2\theta_{13}$
and $\delta$ found in \cite{Capozzi:2017ipn} 
we get for $\Delta m^2_{31(2)} > 0$ 
($\Delta m^2_{31(2)} < 0$):  
$J_{\rm CP}  \approx -\; 0.031$ 
($J_{\rm CP}  \approx -\; 0.026$).
Thus, if the indication that $\delta \approx 3\pi/2$
is confirmed by future more precise data, the CP-violating
effects in neutrino 
oscillations would be relatively large, 
provided the factor $F^{\rm vac}_{\rm osc}$ 
does not lead to a suppression of the asymmetries  $A^{(\alpha,\beta)}_{\rm CP}$. 
Such a suppression would not occur 
if, under the conditions 
of a given experiment, both neutrino mass squared 
differences $\Delta m^2_{21}$ and $\Delta m^2_{31(32)}$
are ``operative'', i.e., 
if neither $\sin(\Delta m^2_{21}L/(2E))\approx 0$ nor 
$\sin(\Delta m^2_{31(32)}L/(2E)) \approx 0$.

 The search for  CP-violating effects from the Dirac phase
in neutrino oscillations 
is one of the principal goals of the future 
experimental studies in neutrino physics
(see, e.g., 
\cite{Adamson:2014vgd, Adamson:2016tbq, Vahle:2016,Abe:2015zbg,Abe:2016ero,Acciarri:2015uup,Acciarri:2016crz,Acciarri:2016ooe,T2K:Nu2016}).
In order for the CP-violating effects in neutrino oscillations 
to be observable, both $\sin(\Delta m^2_{31}L/(2E))$ and
$\sin(\Delta m^2_{21}L/(2E))$ should be sufficiently large.
In the case of  $\sin(\Delta m^2_{31}L/(2E))$, for instance, 
this requires that, say, $\Delta m^2_{31}L/(2E)\approx 1$.
The future experiments on CP violation in 
neutrino oscillations are planned to be performed 
with accelerator $\nu_{\mu}$ and $\bar{\nu}_{\mu}$ beams 
with energies from $\sim 0.7$ GeV to a few GeV. Taking as an 
instructive example $E = 1$ GeV and using the bfv of 
$\Delta m^2_{31} = 2.56\times 10^{-3}~{\rm eV^2}$, 
one has  $\Delta m^2_{31}L/(2E)\approx 1$
for $L \approx 1000$ km. 
Thus, the possibility to observe CP violation in
neutrino oscillations requires the experiments 
to have  long baselines. 
The MINOS, T2K and NO$\nu$A  experiments, for example,  
which provide data on $\nu_{\mu}$ oscillations 
(see, e.g., \cite{PDG2016} and references therein), 
have baselines of approximately 735 km, 295 km and 810 km, 
respectively. 
The planned DUNE \cite{Acciarri:2015uup,Acciarri:2016crz,Acciarri:2016ooe}
and T2HK \cite{Abe:2011ts,Abe:2015zbg,Abe:2016ero} experiments,
which are designed to search for CP violation effects in 
neutrino oscillations, will have  baselines of 1300 km and 295 km, 
respectively. 

Thus, in  MINOS, T2K, NO$\nu$A and in the 
future planned experiments DUNE 
and T2HK, 
the baselines are such that the neutrinos travel 
relatively long distances in the matter of the 
Earth mantle. As is well known, the pattern of neutrino 
oscillations can be changed significantly by 
the presence of matter \cite{Wolfenstein:1977ue,Mikheev:1986gs} 
(see also \cite{Barger:1980tf})
due to the coherent (forward) scattering of neutrinos 
on the ``background'' of electrons ($e^-$), protons ($p$) 
and neutrons ($n$) present in matter.
The scattering generates an effective potential $V_{\rm eff}$ 
in the neutrino Hamiltonian: $H = H_{\rm vac} + V_{\rm eff}$, which
modifies the vacuum lepton mixing angles,
since the eigenstates and eigenvalues of $H_{\rm vac}$ and of 
$H$  differ. This 
leads to different oscillation 
probabilities compared with 
oscillations in vacuum. 
The matter in the Earth (and the Sun) is not charge 
conjugation (C-) symmetric, since
it contains only $e^-$, $p$ and $n$
but does not contain their antiparticles.  
As a consequence, the oscillations 
taking place in the Earth 
are neither CP- nor CPT-invariant \cite{Langacker:1986jv}. 
This complicates the studies of
CP violation due to the Dirac phase $\delta$ 
in long baseline neutrino oscillation experiments.

In the constant
density approximation and keeping terms
up to second order in the two small
parameters $|\Delta m^2_{21}/\Delta m^2_{31}|\ll 1$
and $\sin^2\theta_{13} \ll 1$,
 the expression for the
$\nu_\mu \rightarrow \nu_{e}$ oscillation 
probability in the three-neutrino mixing scheme and
for neutrinos crossing the Earth mantle,
has the  form\cite{Freund:2001pn} (see also \cite{Cervera:2000kp})

\begin{equation}
\label{Earthemu}
P^{3\nu~\text{man}}_{m}(\nu_{\mu} \rightarrow \nu_{e}) \approx
P_0 + P_{\sin\delta} + P_{\cos\delta} + P_3\,.
\end{equation}
%

\noindent  \\In the previous expression we define  

\begin{eqnarray}
\label{P0P3}
\begin{split}
& P_0 = \sin^2\theta_{23}\,\,\frac{\sin^22\theta_{13}}{(A - 1)^2}\,
\sin^2[(A-1)\Delta]\,,\\
& P_{3} = r^2\, \cos^2\theta_{23}\,
\frac{\sin^22\theta_{12}}{A^2}\,\sin^2(A\Delta)\,,\\[0.25cm]
\label{Psind}
& P_{\sin\delta} = -\,r \,\, \frac{8\,J_{\rm CP}}{A(1-A)}\,
\sin\Delta\,\sin ( A\Delta) \, \sin \left[(1-A)\Delta \right]\,,\\[0.25cm]
\label{Pcosd}
& P_{\cos\delta} = r \,\, \frac{8\,J_{\rm CP}\, \cot\delta}{A(1-A)}\,
\cos\Delta\,\sin (A\Delta)\, \sin \left[(1-A)\Delta 
\right]\,,
\end{split}
\end{eqnarray}
%
\noindent where
\begin{eqnarray}
r = \frac{\Delta m^2_{21}}{\Delta m^2_{31}}\,,\quad
~\Delta = \frac{\Delta m^2_{31}\,L}{4E}\,,\quad~
A = \sqrt{2}\,G_{\rm F}\,N^{\rm man}_{e}\frac{2E}{\Delta m^2_{31}}\,,
\label{aDA}
\end{eqnarray}
%
$G_{\rm F}$ is the Fermi constant and  $N^{\rm man}_{e}$ denotes the electron number density of the Earth mantle.
The Earth matter effects on the oscillations 
are accounted for by the quantity $A$.
The mean electron number density
in the Earth mantle 
is $\bar{N}_{e}^{\rm man}\approx 1.5~{\rm cm^{-3}}~N_A$ \cite{Dziewonski:1981xy},
$N_A$ being Avogadro's number. 
$N^{\rm man}_e$ varies little around the indicated mean value along 
the trajectories of the neutrinos
in the Earth mantle corresponding to
the experiments under discussion.
Thus, as far as the calculation of neutrino oscillation
probabilities is concerned, the constant density 
approximation $N_{e}^{\rm man} = \tilde{N}_{e}^{\rm man}$, where
$\tilde{N}_{e}^{\rm man}$ is the (constant) mean density along the
given neutrino path in the Earth,
was shown to be sufficiently accurate
\cite{Krastev:1988yu,Petcov:1998su,Agarwalla:2013tza}.
The 
expression for the probability of 
$\bar{\nu}_{\mu} \rightarrow \bar{\nu}_{e}$
oscillation can be obtained
formally from that for $P^{3\nu~\text{man}}_{m}(\nu_{\mu} \rightarrow \nu_{e})$
by making the changes $A\rightarrow -A$ and $J_{\rm CP}\rightarrow - J_{\rm CP}$,
with $J_{\rm CP}\cot\delta$
remaining unchanged.\footnote{For a detailed discussion of the conditions of 
validity of the analytic expression for 
$P^{3\nu~\text{man}}_{m}(\nu_{\mu} \rightarrow \nu_{e})$
quoted above see  \cite{Freund:2001pn}.} If the Dirac phase
in the PMNS mixing matrix $U_\nu$ has a CP-conserving value, 
we would have $P_{\sin\delta} = 0$.
However, we would still have that
$P^{3\nu~\text{man}}_{m}(\nu_{\mu} \rightarrow \nu_{e})$ and 
$P^{3\nu~\text{man}}_{m}(\bar{\nu}_{\mu} \rightarrow \bar{\nu}_{e})$ are unequal
due to the Earth matter effects.
It is possible, in principle, to 
experimentally
disentangle the effects of the Earth matter
and of $J_{\rm CP}$ in the asymmetries for neutrinos crossing the Earth mantle,  $A^{(e, \mu)~\text{man}}_{\rm CP}$,
by studying the energy dependence of 
$P^{3\nu~\text{man}}_{m}(\nu_{\mu} \rightarrow \nu_{e})$
and  $P^{3\nu~\text{man}}_{m}(\bar{\nu}_{\mu} \rightarrow \bar{\nu}_{e})$.
This will allow to obtain direct information 
about Dirac CP violation in the lepton sector 
and to measure the Dirac phase $\delta$.
In the vacuum limit $N^{\rm man}_e = 0$ ($A = 0$)
we have $A^{(e, \mu)~\text{man}}_{\rm CP} = A^{(e, \mu)}_{\rm CP}$, see \eref{CPTAsym}, and only the term 
$P_{\sin\delta}$ contributes to  $A^{(e, \mu)}_{\rm CP}$. 

 The expressions for the probabilities 
$P^{3\nu~\text{man}}_{m}(\nu_{\mu} \rightarrow \nu_{e})$
and  $P^{3\nu~\text{man}}_{m}(\bar{\nu}_{\mu} \rightarrow \bar{\nu}_{e})$
can be used in the interpretation of the 
results of MINOS, T2K,  NO$\nu$A, and of the 
future planned DUNE and T2HK  experiments. For a 
discussion of the sensitivity of these experiments to 
$\delta$ see, e.g.,  
\cite{Adamson:2014vgd,Adamson:2016tbq, Vahle:2016,Acciarri:2015uup,Acciarri:2016crz,Acciarri:2016ooe,Abe:2015zbg,Abe:2016ero,T2K:Nu2016}. 
If indeed $\delta \approx 3\pi/2$, the DUNE and T2HK  experiments 
are foreseen to establish the existence of leptonic Dirac CP violation 
at $\sim 5\, \sigma$.

%
\subsection{Majorana  phases and neutrinoless double beta decay}
%

If neutrinos are massive Majorana particles, the lepton mixing matrix contains 
two additional Majorana phases  \cite{Bilenky:1980cx}, $\alpha$ and  $\beta$, see \eref{VP}.
However, getting experimental information about these  
 CP-violating phases  is a remarkably difficult problem 
\cite{Bilenky:2001rz,Pascoli:2001by,Barger:2002vy,Pascoli:2005zb,deGouvea:2002gf}.
In fact, the flavor neutrino oscillation probabilities
are insensitive to the CP phases
$\alpha$ and $\beta$.   
The Majorana phases play an important role in processes 
characteristic of the Majorana nature of massive neutrinos.

The massive Majorana neutrinos 
can mediate processes in which the 
total lepton number changes by two units, 
$|\Delta L|= 2$, such as 
$K^+ \rightarrow \pi^- + \mu^+ + \mu^+$ and
$e^- +(A,Z) \rightarrow e^+ + (A,Z-2)$.
The rates of these processes are 
typically proportional to the factor 
$(m_j/M(|\Delta L|= 2))^2$, $M(|\Delta L|= 2)$ 
being the characteristic mass scale of 
the given process, and thus are typically 
extremely small. The experimental searches for neutrinoless double beta
($\betabeta$) decay, 
$(A,Z) \rightarrow (A,Z+2) + e^- + e^-$,
of even-even nuclei, e.g., $^{48}$Ca, $^{76}$Ge, 
$^{82}$Se, $^{100}$Mo, $^{116}$Cd, $^{130}$Te, 
$^{136}$Xe, $^{150}$Nd, 
are unique in reaching a sensitivity
that might allow to observe this process, 
if it is triggered by the exchange of the light 
neutrinos $\nu_j$ or new physics beyond the SM (see, e.g., 
\cite{Petcov:2013poa,DellOro:2016tmg,Rodejohann:2012xd, Pas:2015eia}).
The half-life $T_{1/2}^{0\nu}$ of an isotope decaying via this process, induced by the exchange of virtual $\nu_{1,2,3}$, takes the form (see, e.g., \cite{Vergados:2016hso})

\begin{equation}
	\frac{1}{T_{1/2}^{0\nu}} \; = \; G^{0\nu} \left|\mathcal{M}(A,Z) \right|^2 \frac{\left|m_{\beta\beta} \right|^2}{m_e}\,,
\end{equation}
where $\mathcal{M}(A,Z)$ denotes the nuclear matrix element (NME) for a $|\Delta L|=2$ transition, $G^{0\nu}$ is a phase space factor, $m_e$ the electron mass and $|m_{\beta\beta}|$ the  $\betabeta$ decay effective Majorana mass.
The latter
contains all the dependence of $T_{1/2}^{0\nu}$ 
on the lepton mixing parameters. 
We have 

\begin{equation}
\left|m_{\beta\beta}\right|=\left| m_1 \,c_{12}^2\, c_{13}^2 
\,+\, m_2 \,s_{12}^2\,c_{13}^2 \,e^{2 i\alpha}
 \,+\, m_3 \,s_{13}^2\, e^{2i \beta^\prime} \right|\,,
\label{effmass2}
\end{equation}
%
with  $\beta^\prime\equiv\beta - \delta$. 
For the  NH,  IH
and QD neutrino mass spectra, 
$\left|m_{\beta\beta}\right|$ is given by

\begin{equation}
\begin{split}
\text{NH}\,, \quad\left|m_{\beta\beta}\right|\;\approx\; & \left|\sqrt{\Delta m^2_{21}}~s^2_{12}
+ \sqrt{\Delta m^2_{31}}~s^2_{13}e^{2i(\beta^\prime-\alpha)}\right|\,,~~\\ 
\text{IH}\,,\quad \left|m_{\beta\beta}\right| \;\approx\;& \sqrt{|\Delta m^2_{32}|}~   
\left|c^2_{12} + s^2_{12}~e^{2i\alpha}\right|\,,~\\
\text{QD}\,, \quad\left|m_{\beta\beta}\right| \;\approx\; &  m_{\rm QD}~\left|c^2_{12}
 + s^2_{12}~e^{2 i \alpha}\right|\,.~
\end{split}
\end{equation}
Obviously, $\left|m_{\beta\beta}\right|$  strongly  depends
on the Majorana phases:
the CP-conserving values of 
$\alpha$=$0,\pi/2$ \cite{Wolfenstein:1981rk,Bilenky:1984fg,Kayser:1984ge}, 
for instance, determine the range of 
possible values of $\left|m_{\beta\beta}\right|$ in the 
cases of IH and QD spectrum. 

 Using the $3\, \sigma$ ranges of the allowed values of 
the neutrino oscillation parameters 
quoted in 
 \eref{deltasolvalues},
one finds that:\\
i) $0.78\times 10^{-3}~{\rm eV} \ltap \left|m_{\beta\beta}\right|
\ltap 4.32\times 10^{-3}$ eV in the case of NH spectrum;\\
ii) $1.4\times 10^{-2}~{\rm eV} \ltap \left|m_{\beta\beta}\right| 
\ltap 5.1\times 10^{-2}$ eV in the case of IH spectrum;\\
iii) $m_{\rm QD}/3 \ltap \left|m_{\beta\beta}\right| \ltap m_{\rm QD}$,
$m_{\rm QD} \gtap 0.10$ eV, in the case of QD spectrum.\\ 
The difference in the ranges of 
$\left|m_{\beta\beta}\right| $ in these three cases opens
up the possibility to get information about
the type of neutrino mass spectrum from a
measurement of $\left|m_{\beta\beta}\right|$~\cite{Pascoli:2002xq}.
The  main features of the 
predictions for $\left|m_{\beta\beta}\right|$ are  illustrated 
\begin{figure}[t!]
\centering
\includegraphics[width=9.5cm,height=7.0cm,clip=]{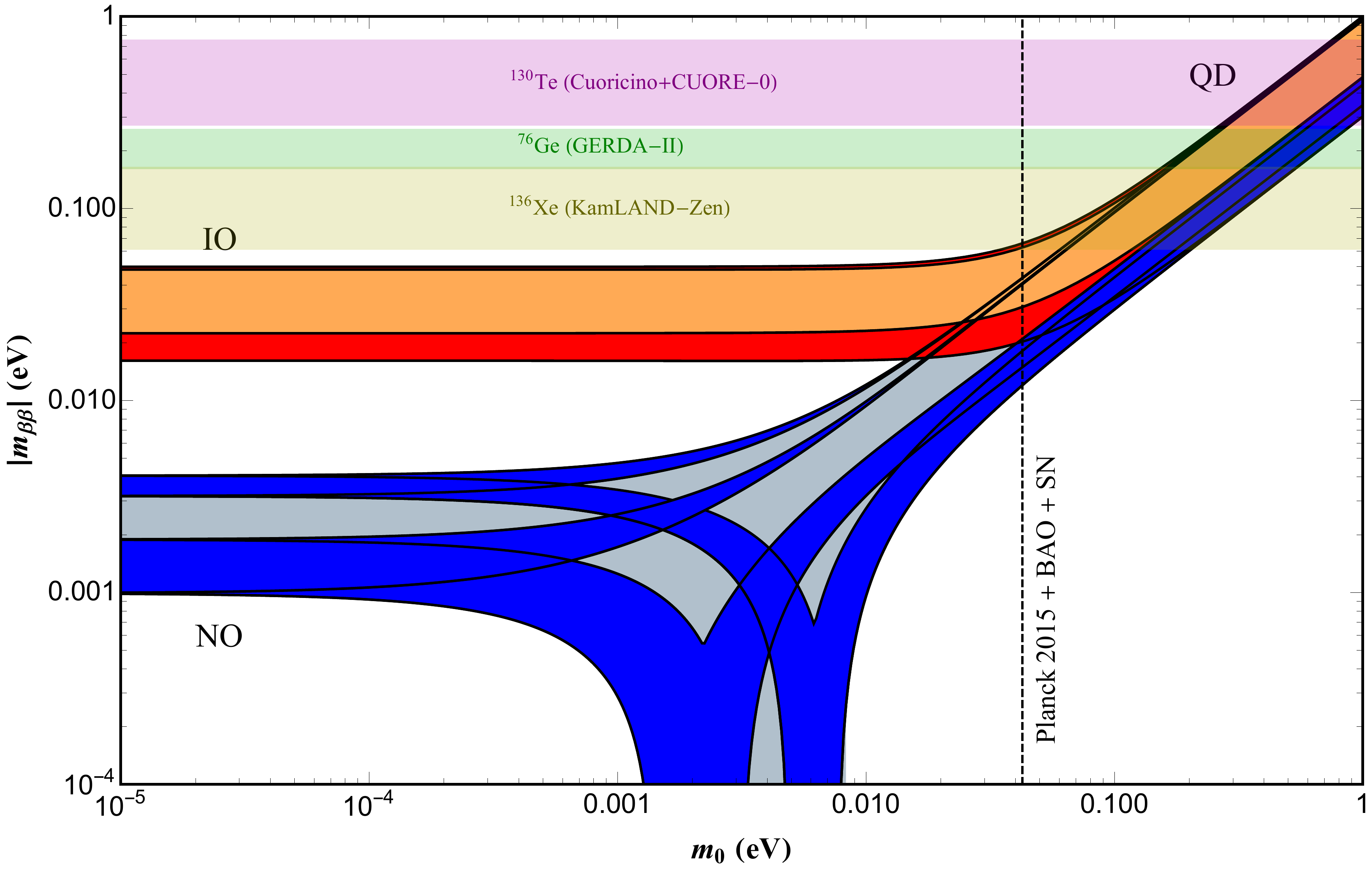}
\caption{
The effective Majorana mass $\left|m_{\beta\beta}\right|$
as a function of the lightest neutrino mass $m_0$. 
The figure is obtained using 
the bfvs and the $2\, \sigma$ 
ranges of allowed values 
of  $\Delta m^2_{21}$, $\sin^2\theta_{12}$, and
$|\Delta m^2_{31}| \approx |\Delta m^2_{32}|$, $\sin^2\theta_{13}$
from \protect\cite{Capozzi:2013csa}.
The CP phases $\alpha$ and $\beta^\prime$
are varied in the interval [0,$\pi$). 
The predictions for the NO, IO and QD mass
spectra are indicated.
In the light-blue and orange regions at least one of
the phases $\alpha,\beta^\prime$
and $(\beta^\prime - \alpha)$ has a CP-violating value, while the 
blue and red areas correspond to $\alpha$ and $\beta^\prime$
possessing CP-conserving values. The 90\% CL  exclusion limits on $\left|m_{\beta\beta}\right|$ from searches for $\betabeta$ decay and the 95\% CL upper bound on $m_0$
from cosmological data are indicated.}
\label{Fig1}
\end{figure}
%
\noindent  
in \fref{Fig1}, 
where $\left|m_{\beta\beta}\right|$ is shown as a function of the lightest 
neutrino mass $m_0$.

 The predictions for $\left|m_{\beta\beta}\right|$ in the cases of the NO, IO and QD 
neutrino mass spectra can be drastically modified  
by the existence of lepton number violating $|\Delta L| = 2$
new physics beyond that predicted by the SM: 
eV or GeV to TeV scale RH neutrinos, etc. 
(see, e.g., \cite{Bilenky:2001xq,Barry:2011wb,Girardi:2013zra,Blennow:2010th,Lopez-Pavon:2015cga,Deppisch:2012nb,Deppisch:2015qwa}). 
There is a potential synergy between the searches for 
$\betabeta$ decay and the searches for neutrino-related 
$|\Delta L| = 2$ beyond the SM physics at LHC \cite{Deppisch:2015qwa}.

The best lower limit on the half-life of $^{76}$Ge,
$T_{1/2}^{0\nu}(^{76}\text{Ge}) > 5.2\times 10^{25}$~yr (at 90\%~CL),  
was found in the GERDA-II $^{76}$Ge 
experiment \cite{GERDA:Nu2016}. 
This lower limit on $T_{1/2}^{0\nu}(^{76}\text{Ge})$ corresponds to 
the upper limit $\left|m_{\beta\beta}\right| < (0.16 \div 0.26)$ eV,
in which the estimated uncertainties in the NME 
are accounted for. 
 Two experiments, NEMO-3 \cite{NEMO3:Nu2016} with $^{100}$Mo 
and CUORICINO+CUORE-0 \cite{CUORI:Nu2016}  
with $^{130}$Te, obtained the limits (at 90\% CL): 
 $T_{1/2}^{0\nu}(^{100}\text{Mo}) > 1.1\times 10^{24}$~yr  
corresponding to
$\left|m_{\beta\beta}\right| < (0.3 \div 0.6)$ eV \cite{NEMO3:Nu2016}
with the estimated NME uncertainties taken into account,
and $T_{1/2}^{0\nu}(^{130}\text{Te}) > 4.4\times 10^{24}$~yr 
\cite{CUORI:Nu2016}. 
 The best lower limit on 
 the $\betabeta$ decay half-life of $^{136}$Xe 
 was obtained by the 
KamLAND-Zen Collaboration \cite{KamLAND-Zen:2016pfg}: 
$T_{1/2}^{0\nu}(^{136}\text{Xe}) > 1.07\times 10^{26}$~yr,
corresponding to  $\left|m_{\beta\beta}\right|< (0.061 \div 0.165)$ eV
(at 90\%~CL). 
The EXO Collaboration \cite{Auger:2012ar} reported the 
lower limit:  
$T_{1/2}^{0\nu}(^{136}\text{Xe}) > 1.6\times 10^{25}$~yr 
(at 90\%~CL).

 Most importantly, a large number of 
experiments of a new generation aim at
sensitivities to $\left|m_{\beta\beta}\right| \sim (0.01\div 0.05)$ eV (see, e.g., 
\cite{DellOro:2016tmg}): CUORE
($^{130}$Te), SNO+ ($^{130}\text{Te}$),
GERDA ($^{76}$Ge), MAJORANA ($^{76}$Ge), LEGEND ($^{76}$Ge),
SuperNEMO ($^{82}$Se, $^{150}$Nd), 
 KamLAND-Zen ($^{136}$Xe), 
EXO  and nEXO ($^{136}$Xe), PANDAX-III  ($^{136}$Xe),
NEXT ($^{136}$Xe), AMoRE ($^{100}$Mo), MOON ($^{100}$Mo),  
CANDLES ($^{48}$Ca), 
XMASS  ($^{136}$Xe),
DCBA  ($^{82}$Se, $^{150}$Nd),
ZICOS  ($^{96}$Zr), etc.~The GERDA-II and KamLAND-Zen 
experiments have already provided  
the best lower limits on the $\betabeta$ decay
half-lives of $^{76}$Ge and $^{136}$Xe. 
The experiments listed above
aim to probe the QD and IO ranges of $\left|m_{\beta\beta}\right|$.
If  $\betabeta$ decay will be 
observed in these experiments, 
the measurement of the $\betabeta$ decay 
half-life might allow to obtain constraints on 
the Majorana phase $\alpha$, 
see \cite{Bilenky:2001rz,Pascoli:2001by,Bilenky:1996cb,Faessler:2008xj}.

Proving that the CP symmetry is violated 
in the lepton sector 
due to the Majorana 
phases $\alpha$ and $\beta^{(\prime)}$ is remarkably challenging 
\cite{Pascoli:2001by,Barger:2002vy,Pascoli:2005zb}:
it requires quite accurate measurements 
of $\left|m_{\beta\beta}\right|$ (and of $m_{\rm QD}$ for QD spectrum), 
and holds only for a limited range of 
values of the relevant parameters.
For $\sin^2\theta_{12}$=0.31
in the case of QD spectrum, for example,
establishing at 95\% CL 
that the relevant phase 
$\alpha$ possesses a CP-violating value
requires  \cite{Pascoli:2005zb}
a relative error on the measured value of 
$\left|m_{\beta\beta}\right|$ and $m_0$ smaller than 15\%,
a theoretical uncertainty
$F{\small \ltap}$1.5 in 
 the corresponding NME, 
and a value of 
$\alpha$ typically within the ranges 
of $(\pi/8 \div 3\pi/8)$ and
$(5\pi/8 \div 7\pi/8)$.

 Obtaining quantitative information on
the lepton mixing parameters from a measurement of
$\betabeta$ decay half-life would be impossible 
without sufficiently precise knowledge of the 
corresponding NME of the process.\footnote{For discussions of the current status of the 
calculations of the NMEs for  $\betabeta$ decay see, e.g., 
\cite{Vergados:2016hso}. 
A possible test of the NME calculations is 
suggested in \cite{Pascoli:2001by} and 
is discussed in greater detail in \cite{Bilenky:2004um}.  
}  
At present the variation of the values
of the different $\betabeta$ decay NMEs 
calculated using the various currently employed methods 
is typically by a factor of two to three.
The observation of $\betabeta$ decay of 
one nucleus is likely to lead to the searches and 
observation of the decay of other nuclei.
The data on the  $\betabeta$ decay of 
several nuclei might help to 
solve the problem of sufficiently precise 
calculation of the $\betabeta$ decay NMEs \cite{Pascoli:2001by}. 

An additional source of uncertainty 
is the effective value of the axial-vector 
coupling constant $g_A$ in $\betabeta$ decay. 
This constant is renormalized by nuclear medium effects, which tend to 
quench, i.e., reduce, the vacuum value of $g_A$. 
The problem of the quenching of  $g_A$ arose in connection with the 
efforts to describe theoretically the experimental 
data on the two-neutrino double beta decay.
The physical origin of the quenching is not fully understood, 
and the size of the quenching of $g_A$ in $\betabeta$ decay 
is subject to debates (for further details see, e.g., \cite{Vergados:2016hso}).
The importance of the effective value of $g_A$ in 
$\betabeta$ decay stems from the fact that, to a good approximation, 
the $\betabeta$ decay rate is proportional to the fourth power of 
the effective $g_A$.

  If the future $\betabeta$ decay experiments 
show that $\left|m_{\beta\beta}\right| < 0.01$ eV, both the IO and
the QD spectrum will be ruled out for massive 
Majorana neutrinos. If in addition it is 
established in neutrino oscillation 
experiments that $\Delta m^2_{31(32)} < 0$ 
(IO spectrum), one would be led to conclude that 
either the massive neutrinos $\nu_j$ 
are Dirac fermions, or that 
$\nu_j$ are Majorana particles,
but there are additional contributions to the 
$\betabeta$ decay amplitude which 
interfere destructively 
with that due to the exchange of $\nu_j$. 
The case of more than one mechanism 
generating the $\betabeta$ decay is discussed 
in, e.g., \cite{Faessler:2011qw,Meroni:2012qf},
where the possibility to identify the mechanisms 
inducing the decay is also analyzed.
If, however, $\Delta m^2_{31(32)}$ 
is determined to be positive,
the upper limit $\left|m_{\beta\beta}\right| < 0.01$ eV 
would be perfectly compatible with 
massive Majorana neutrinos
possessing NH mass spectrum, 
or NO spectrum with  partial hierarchy,
and the quest for $\left|m_{\beta\beta}\right|$ 
would still be open \cite{Pascoli:2007qh}.

\section{Type I seesaw mechanism of neutrino mass generation and  leptogenesis}
\label{sec3}

It follows from the neutrino data summarized in the previous 
sections that neutrino masses are much smaller 
than the masses of  the other SM fermions. If we take as an 
indicative upper limit $m_i \ltap 0.5$ eV, we have
 $m_i/m_e\ltap 10^{-6}$.
It is natural to suppose that the remarkable
smallness of neutrino masses is related to the existence of 
a new fundamental mass scale in particle physics, 
and thus to new physics beyond 
that predicted by the SM.

A possible explanation of neutrino masses is provided by the seesaw  
mechanism of neutrino mass generation. 
In the type I seesaw realization \cite{Minkowski:1977sc,Yanagida:1979as,GellMann:1980vs,Mohapatra:1979ia}, the SM is extended with at 
least two heavy RH neutrinos $N_a$, with $P_R N_a = N_a$, 
coupled to the 
SM leptons via the seesaw Lagrangian
\begin{equation}
\mathcal{L} \;=\; \mathcal{L}_{\rm SM} +  \bar{N}_a \, i \slashed\partial \, N_a - \left( \frac 12   \left(M_N\right)_{a b} \bar{N}_a^c N_b + \lambda_{\alpha b} \, \bar{\ell}_\alpha \phi^c N_b + \text{h.c.} \right),
\label{eq:seesawI}
\end{equation}
where $\ell_\alpha \equiv \left( \nu_{L\alpha},\, e_{L\alpha} \right)^T$ is the SM lepton doublet of a given flavor $\alpha=e,\mu,\tau$, $\phi^c\equiv \epsilon\, \phi^\ast$  
($\epsilon_{12}=-1$) is the conjugate Higgs doublet and $N^c_a\equiv C\bar{N}_a^T$, with $P_L N^c_a = N^c_a$.  We denote by 
$\mathcal{N}_k = (U_R)_{ak}^* N_a$ the RH neutrino fields with definite masses $M_k>0$, $U_R$ being the unitary matrix which diagonalizes the Majorana mass matrix $M_N$ in 
Eq.~\eqref{eq:seesawI}.
In the discussion below we 
consider  the case of three RH neutrinos, $a, k =1,2,3$.

After electroweak symmetry breaking, the neutrino Yukawa interactions  
in \eref{eq:seesawI} generate a Dirac neutrino mass term 
$(m_D)_{\alpha b} \,\overline{\nu}_{L\alpha} N_{b}$, 
where $m_D=\lambda\,v/\sqrt{2}$ and $v=246$ GeV
is the vacuum expectation value of the SM Higgs. 
Under the assumption that $v\ll M_k$,  
the heavy RH neutrinos $\mathcal{N}_k$ 
provide at tree-level an effective Majorana mass term for the active neutrino fields $\nu_{L \alpha}$, i.e., $\frac12(M_\nu)_{\alpha\beta}\, \overline{\nu}_{L\alpha} \nu_{L\beta}^c$, with

\begin{equation}
	M_\nu \;\approx\; - \frac 12 \,v^2\,\lambda\, M_N^{-1} \, \lambda^T\,=\, -m_D\,M_N^{-1}\,m_D^T\,=\, U_\nu\, M_\nu^{\rm diag}\, U_\nu^T\,,
\end{equation}
$M_\nu^{\rm diag}$ containing the light neutrino masses $m_i$ as diagonal entries.
Taking indicatively $M_\nu\sim 0.05$ eV and $m_D\sim 200$ GeV, one predicts a seesaw  lepton number violating scale: $M_N\sim M_k\sim 10^{14}$ GeV. 

In the early Universe, these RH neutrinos are produced by scattering processes involving SM particles in thermal equilibrium at temperatures $T\gtrsim M_k$.
Their out-of-equilibrium decays  generate a lepton asymmetry which is partially converted into a non-zero baryon asymmetry by fast sphaleron processes \cite{Kuzmin:1985mm}, a mechanism called
 \textit{thermal leptogenesis} \cite{Fukugita:1986hr}. A pedagogical discussion of this mechanism is given in \cite{leptogenesis:A04}.

 The cosmological  baryon asymmetry $Y_{\Delta B}$  can be expressed in terms 
of the ratio between the net baryon number density and the entropy 
density of the Universe
\begin{equation}
	Y_{\Delta B} \,\equiv \, \left. \frac{n_B - n_{\bar{B}}}{s}\right|_0\,=\,\left(8.65\pm0.09 \right)\times 10^{-11}\,,
\end{equation}
where the quoted bfv~and $1\, \sigma$ error are taken from \cite{Ade:2015xua} 
and the subscript ``0" refers to the present epoch.

If the spectrum of the heavy RH neutrinos $\mathcal{N}_k$ is hierarchical,  
with $M_1\ll M_2\lesssim M_3$, most of the baryon asymmetry is 
produced by the out-of-equilibrium decay of the lightest state, $\mathcal{N}_1$, 
at temperatures $T\lesssim M_1$, provided the corresponding neutrino 
Yukawa couplings  
have sufficiently large
CP-violating phases. For $M_1\gtrsim 10^{13}$ GeV one has within 
the SM (see, e.g., \cite{Davidson:2008bu})
\begin{equation}
	Y_{\Delta B} \approx  -\,1.39 \times 10^{-3} \, \epsilon_1 \, \eta\,, 
\label{YBdef1}
\end{equation}
%
where $10^{-3}\lesssim \eta\lesssim 1$ is an efficiency factor which takes 
into account the washout of the lepton asymmetry \cite{Giudice:2003jh} 
by lepton number violating processes occurring in the
thermal bath, whereas $\epsilon_1$ denotes the total CP asymmetry 
in $\mathcal{N}_1$ decays. 

For a RH neutrino $\mathcal{N}_k$
decaying into a (given) 
lepton (of flavor $\alpha$), the (flavored) CP asymmetry 
$\epsilon_{k(\alpha)}$ is  defined as 
\begin{eqnarray}
	\epsilon_{k\alpha} &\equiv& \frac{\Gamma\left(\mathcal{N}_k\to \phi\,\ell_\alpha\right)\,-\,\Gamma\left(\mathcal{N}_k\to \bar{\phi}\,\bar{\ell}_\alpha\right)}
	{\sum_\beta \left[\Gamma\left(\mathcal{N}_k\to \phi\,\ell_\beta\right)\,+\,\Gamma\left(\mathcal{N}_k\to \bar{\phi}\,\bar{\ell}_\beta\right)\right]}  \nonumber\\
	& = &  \frac{1}{4 \pi v^2\,(\hat{m}_D^\dagger \hat{m}_D)_{kk} } \sum_{j \neq k}\Big\{  \mathrm{Im} \left((\hat{m}_D^\dagger \hat{m}_D)_{kj}  (\hat{m}_D)^\ast_{\alpha k} (\hat{m}_D)_{\alpha j} \right) \, f(M_j/M_k) \nonumber\\
&&+  \,
\mathrm{Im} \left( (\hat{m}_D^\dagger \hat{m}_D)_{jk} (\hat{m}_D)^\ast_{\alpha k} (\hat{m}_D)_{\alpha j}  \right) \, g\left(M_j/M_k \right) \Big\}\,\label{epskagen}
\end{eqnarray}
and
\begin{eqnarray}
\begin{split}
	\epsilon_k & \equiv  \sum\limits_{\alpha=e,\mu,\tau} \epsilon_{k\alpha} \\
	& =   \frac{1}{4 \pi v^2\,(\hat{m}_D^\dagger \hat{m}_D)_{kk} } \sum_{j \neq k} \mathrm{Im} \left((\hat{m}_D^\dagger \hat{m}_D)_{kj}^2 \right)\, f(M_j/M_k)\,,
\end{split}	
\end{eqnarray}
%
where $f(x)$ and $g(x)$ are known loop functions \cite{Covi:1996wh}, namely
\begin{eqnarray}
\begin{split}\label{loopfun}
	f(x) & = x\left[ \frac{1}{1-x^2}+1-(1+x^2)\ln\left(1\,+\,\frac{1}{x^2}\right)\right]\,,\\ 
	g(x) & = \frac{1}{1-x^2}\,.
\end{split}
\end{eqnarray}
%
In \eref{epskagen} $\hat{m}_D=m_D U_R$ 
is the Dirac neutrino mass matrix in the RH neutrino mass basis.
Successful leptogenesis is possible for the lightest neutrino mass
 $m_{0} \lesssim 0.1$ eV \cite{Davidson:2008bu,Fong:2013wr,Blanchet:2012bk}, 
which is compatible with the current constraints 
on  $m_0$. 
In certain cases of IH neutrino mass spectrum
and 
 values of  $5\times 10^{-4}~\text{eV}\lesssim m_3\lesssim 10^{-2}~\text{eV}$
one can have very strong amplification of the  baryon asymmetry produced 
via leptogenesis with respect to the case of  $m_3 = 0$ \cite{Molinaro:2007uv}.

The expression of the baryon asymmetry given in \eref{YBdef1} is modified 
in the case the RH neutrino mass spectrum is not strongly 
hierarchical and/or flavor dynamics becomes relevant in leptogenesis \cite{Barbieri:1999ma,Abada:2006fw,Abada:2006ea,Nardi:2006fx}, 
which is the case if $M_1\lesssim 10^{12}$ GeV 
 (see \cite{leptogenesis:A01} for a detailed discussion of flavor effects in leptogenesis). 
In general, within the \textit{flavored leptogenesis} regime, 
the physical CP phases of the neutrino Yukawa couplings $\lambda_{\alpha b}$ 
are not directly related to the CP phases potentially measurable at 
low energies, i.e. the Dirac phase $\delta$ and  the two Majorana 
phases $\alpha$ and $\beta$ in \eref{VP} and \eref{eq:Vpara} 
(see, e.g., \cite{Branco:2001pq,Rebelo:2002wj,Pascoli:2003uh,Davidson:2007va,Davidson:2008pf}). 

However, low energy CP phases may also play a crucial role 
in having successful leptogenesis in the case when flavor effects are 
relevant \cite{Pascoli:2006ie,Pascoli:2006ci,Molinaro:2007uv,Molinaro:2008rg,Molinaro:2008cw,Antusch:2006gy,Blanchet:2006be}.
In fact, one can show that if $10^{8}~\text{GeV}\lesssim M_1 \lesssim 10^{12}~\text{GeV}$, CP violation necessary for successful leptogenesis can be provided entirely by the CP phases in the lepton mixing matrix.
In particular, under the assumption that the only source of CP violation 
in the seesaw Lagrangian in \eref{eq:seesawI} is given by the Dirac phase 
$\delta$ in the PMNS mixing matrix, the observed value of $Y_{\Delta B}$ can be 
achieved, provided $M_1\lesssim 5\times 10^{11}$ GeV and $\sin\theta_{13}|\sin\delta| \gtrsim 0.09$ ($|J_{\rm CP}|\gtrsim 0.02$) 
\cite{Pascoli:2006ie,Pascoli:2006ci}. 
The predicted lower bound on $\sin\theta_{13}$  ($|J_{\rm CP}|$)
is compatible with current hints of maximal CP violation in neutrino 
oscillations, i.e. $\delta\approx 3\pi/2$. One can also show that even 
if additional sources of CP violation are introduced in the seesaw Lagrangian, 
there are large regions of the parameter space, where the observed 
baryon asymmetry is reproduced, only if we assume non-trivial low energy 
CP-violating Dirac and/or Majorana phases in the lepton mixing matrix 
\cite{Molinaro:2008cw,Molinaro:2008rg}. For example, in the case 
of $5\times 10^{10}~\text{GeV}\lesssim M_1\lesssim 10^{12}~\text{GeV}$, 
IH  light neutrino mass spectrum and $\sin\theta_{13}\cos\delta\lesssim -0.1$ 
a sufficiently large baryon asymmetry is possible, only if the 
requisite CP violation is provided by the non-trivial  Majorana 
phase $\alpha$ in the lepton mixing matrix, independently of  
possible high energy CP phases in the seesaw Lagrangian.

A direct connection between low energy CP violation in the lepton sector 
and the source of CP violation required to have successful leptogenesis 
in the early Universe is possible in
several seesaw scenarios which implement  flavor (and CP)  symmetries  in order to explain the neutrino oscillation data. We will explore  in the next section the implications for leptogenesis in some representative scenarios \cite{Mohapatra:2004hta,Mohapatra:2005ra,Hagedorn:2009jy,Mohapatra:2015gwa,Hagedorn:2016lva}.

\section{Flavor (and CP) symmetries for leptogenesis}
\label{sec4}

In this section we discuss leptogenesis in scenarios with a flavor (and CP) symmetry that can constrain 
the mixing angles (and CP phases) in the high and low energy sector. We focus on neutrinos being Majorana particles and
on the type I seesaw mechanism with three RH neutrinos as the source of light neutrino masses. The decay of
RH neutrinos $\mathcal{N}_k$ leads to the generation of the CP asymmetries $\epsilon_{k (\alpha)}$ that are converted into the cosmological baryon asymmetry $Y_{\Delta B}$.

It is well-known that flavor (and CP) symmetries are a powerful tool for explaining the measured values of the 
mixing angles in the lepton sector and for making predictions for the yet-to-be-measured Dirac and Majorana phases.
The two central assumptions~\cite{Lam:2007qc,Blum:2007jz,Hernandez:2012ra} are that the three generations of leptons form a (irreducible) triplet,  $\ell_\alpha \sim {\bf 3}$, of a flavor symmetry $G_f$
 (and CP) and that such symmetry is not broken arbitrarily, but rather to non-trivial residual symmetries $G_e$
in the charged lepton and $G_\nu$ in the neutrino sector, respectively. 

We present examples of different scenarios: the baryon asymmetry $Y_{\Delta B}$ is either generated 
via unflavored~\cite{Mohapatra:2004hta,Grimus:2003sf,Mohapatra:2005ra,Jenkins:2008rb,Hagedorn:2009jy,Bertuzzo:2009im,AristizabalSierra:2009ex,Grimus:2003yn,Hagedorn:2016lva} or via flavored leptogenesis~\cite{Mohapatra:2015gwa,Bertuzzo:2009im,Chen:2016ptr,Gehrlein:2015dxa}; these scenarios possess either a flavor symmetry 
$G_f$ only \cite{Mohapatra:2004hta,Grimus:2003sf,Mohapatra:2005ra,Jenkins:2008rb,Hagedorn:2009jy,Bertuzzo:2009im,AristizabalSierra:2009ex}, or $G_f$ together 
with a CP symmetry \cite{Grimus:2003yn,Mohapatra:2015gwa,Hagedorn:2016lva,Chen:2016ptr} that in general acts non-trivially on flavor space; and the generation of non-vanishing
$Y_{\Delta B}$ might \cite{Jenkins:2008rb,Hagedorn:2009jy,Bertuzzo:2009im,AristizabalSierra:2009ex,Hagedorn:2016lva,Gehrlein:2015dxa} or might not \cite{Mohapatra:2004hta,Grimus:2003sf,Mohapatra:2005ra,Mohapatra:2015gwa,Chen:2016ptr} require corrections to the mentioned breaking scheme of flavor (and CP) symmetries. 
If it does, the smallness of these corrections can also explain the smallness of $Y_{\Delta B}$.

Moreover, we briefly highlight the impact of the constraints on Majorana phases on the effective Majorana mass $\left| m_{\beta\beta}\right|$, as discussed in \cite{Ding:2014ora,Mohapatra:2015gwa,Hagedorn:2016lva,King:2014rwa,Petcov:2014laa}.

\mathversion{bold}
\subsection{Impact of $G_f$ (and  CP) on lepton mixing parameters}\label{sec4.1}
\mathversion{normal}

As shown in \cite{Fukuyama:1997ky,Mohapatra:1998ka}, maximal atmospheric mixing and $\theta_{13} = 0$, the latter by now excluded experimentally to high degree, 
can be obtained from a light neutrino mass matrix $M_\nu$ that is invariant under the exchange of its second and third rows and columns
in the charged lepton mass basis ($M_e$ is diagonal). Thus, the presence of different residual flavor symmetries, i.e.~the $\mu-\tau$ exchange symmetry $G_\nu=Z_2^{\mu\tau}$ in the neutrino sector, represented by the three-by-three matrix $P^{\mu\tau}$ that exchanges the second and third (lepton) generation, and the 
discrete charged lepton flavor symmetry $G_e=Z_3$ (or in its continuous version $G_e=U(1)$), leads to the prediction
of two of the three lepton mixing angles. The maximal size of $G_\nu$ in the case of three generations of Majorana neutrinos is $G_\nu=Z_2 \times Z_2$, as shown in \cite{Lam:2006wm}, and, if it is realized,
also the solar mixing angle $\theta_{12}$ can be fixed. The most prominent example is tribimaximal (TB) lepton mixing with $\sin^2 \theta_{12}=1/3$, proposed in \cite{Harrison:2002er,Xing:2002sw}.
This mixing pattern can be obtained from the flavor symmetries
$G_f=A_4$, see \cite{Altarelli:2005yp,deMedeirosVarzielas:2005qg,Altarelli:2005yx,He:2006dk}, and $G_f=S_4$, see \cite{Lam:2008rs}, and their peculiar breaking to $G_e=Z_3$ in the charged lepton and to $G_\nu=Z_2 \times Z_2$ in the neutrino sector.\footnote{In the case of $G_f=A_4$, $G_\nu$ is partly contained in $G_f$ and partly accidental, whereas $G_\nu$ is a subgroup of $G_f$
 in the case of $G_f=S_4$.} 
 Much effort has been made in order to understand the observed values of the lepton mixing
angles with the help of flavor symmetries $G_f$ and their breaking patterns \cite{deAdelhartToorop:2011re,King:2013vna,Hagedorn:2013nra,Fonseca:2014koa} (for reviews see \cite{Altarelli:2010gt,King:2014nza,Ishimori:2010au,Grimus:2011fk}). In the most predictive version of 
this approach, in which all lepton generations are distinguished by the residual symmetries and the maximal symmetry $G_\nu=Z_2 \times Z_2$ is chosen, 
the charged lepton mass matrix $M_e$ and the light neutrino mass matrix $M_\nu$ are constrained by 
\begin{equation}
g_e ({\bf 3})^\dagger \, M_e \, M_e^\dagger \, g_e ({\bf 3}) =  M_e \, M_e^\dagger \;\; \mbox{and} \;\; g_{\nu, i} ({\bf 3})^\dagger \, M_\nu \, g_{\nu, i}({\bf 3})^\ast = M_\nu \; , \; i=1, \, 2 \; ,
\end{equation}

\noindent with $g_e ({\bf 3})$ and $g_{\nu, i} ({\bf 3})$ being the generators of $G_e$ and $G_\nu$, respectively, in the (irreducible) three-dimensional representation ${\bf 3}$. 
 In this case the numerical values of all three lepton mixing angles can be explained and the Dirac phase $\delta$ is predicted. The main
result of surveys \cite{deAdelhartToorop:2011re,King:2013vna,Hagedorn:2013nra,Fonseca:2014koa} of flavor symmetries $G_f$ being subgroups of $SU(3)$ or $U(3)$ is the following: all mixing patterns, that are in accordance with the experimental data at the $3\, \sigma$ level or better, 
predict a lower bound on the solar mixing angle, $\sin^2 \theta_{12} \gtrsim 1/3$, as well as CP-conserving values of $\delta$. In less constrained approaches
also values of $\delta$ different from $0$ and $\pi$ can be obtained \cite{Hernandez:2012ra,Petcov:2014laa,Girardi:2014faa,Ballett:2014dua,Girardi:2015vha,Girardi:2015rwa,Girardi:2016zwz}.

In view of the recent experimental indications of $\delta$ close to $3 \, \pi/2$ different ways to obtain lepton mixing patterns with such feature  
have been explored. One very promising approach involves in addition to $G_f$ a CP symmetry \cite{Feruglio:2012cw} (see also \cite{Holthausen:2012dk,Chen:2014tpa}). The latter acts in general
non-trivially on flavor space \cite{Grimus:1995zi} and can be represented in the three-dimensional representation ${\bf 3}$, under the assumption that CP is an involution, by a CP transformation $X ({\bf 3})$ which is
 a unitary and symmetric three-by-three matrix. A prominent example of this type of CP symmetry is the $\mu-\tau$ reflection symmetry \cite{Harrison:2002kp} where 
  $X ({\bf 3})=P^{\mu\tau}$. Again, we
assume that non-trivial residual symmetries are preserved in the charged lepton and neutrino sector, respectively.
 In the charged lepton mass basis imposing $X ({\bf 3})= P^{\mu\tau}$ alone
on the light neutrino mass matrix $M_\nu$, i.e.~$P^{\mu\tau} \, M_\nu^\ast \, P^{\mu\tau}=M_\nu$, is sufficient for achieving maximal atmospheric mixing, a maximal CP phase $\delta$ ($\delta=\pi/2$ or $\delta= 3\, \pi/2$) and 
CP-conserving Majorana phases, while the values of $\theta_{13}$ and $\theta_{12}$ are not fixed \cite{Harrison:2002kp}. 
The number of free parameters in  lepton mixing can be
reduced to a single real one, if $M_\nu$ is additionally invariant under a residual flavor symmetry $Z_2$ \cite{Feruglio:2012cw}. Compared to the previous approach, 
 $G_e$ is chosen in the same way, i.e.~as a subgroup of $G_f$, and the form of $G_\nu$ is changed from $G_\nu =Z_2 \times Z_2$ to the direct product of a $Z_2$ subgroup of $G_f$
and the CP symmetry, $G_\nu=Z_2 \times CP$. Consequently, the constraints on $M_\nu$ read in general 
\begin{equation}
g_\nu ({\bf 3})^\dagger \, M_\nu \, g_\nu^\ast ({\bf 3}) = M_\nu \;\; \mbox{and} \;\; X ({\bf 3}) \, M_\nu^\ast \, X ({\bf 3}) = M_\nu \;.
\end{equation}
 \noindent One free real parameter $\theta$ is introduced in the 
lepton mixing pattern, since the
residual flavor symmetry in the neutrino sector is reduced from $Z_2 \times Z_2$ to $Z_2$ only. The predictive power regarding CP phases
is, however, enhanced, as CP is involved as symmetry, and both Majorana phases are also predicted. Hence, all lepton mixing parameters only depend on $\theta$, 
that has to be adjusted to a particular value in order to 
obtain lepton mixing angles in accordance with the experimental data. 
By now many flavor symmetries $G_f$ have been combined with CP
and several interesting mixing patterns have been discussed in 
\cite{Hagedorn:2014wha,Ding:2014ora,DiIura:2015kfa,Li:2015jxa,
Ballett:2015wia,Rong:2016cpk}.

 Corrections to this picture such as those arising from renormalization
group running and from perturbations of the symmetry breaking pattern 
strongly depend on the explicit model, e.g.~on the light 
neutrino mass spectrum \cite{Babu:1993qv,Antusch:2001ck}, whether the 
theory is supersymmetric 
or not \cite{Altarelli:2005yp,He:2006dk,Altarelli:2005yx,deMedeirosVarzielas:2005qg},
which auxiliary symmetries are present
in the model. They are assumed to have a small impact on 
lepton mixing parameters.

Throughout the analyses focussing on lepton mixing patterns
the mechanism that generates light neutrino masses is not specified \cite{Lam:2007qc,Blum:2007jz,Hernandez:2012ra,Feruglio:2012cw}. 
However, these approaches can be combined with various
 generation mechanisms for light neutrino masses. In particular, 
they can be implemented in scenarios with the type I seesaw
mechanism. The RH neutrinos $N_a$ in Eq.~\eqref{eq:seesawI} are assigned to the same 
triplet ${\bf 3}$ as the lepton generations $\ell_\alpha$. Two different mass matrices, 
the Dirac neutrino mass matrix $m_D$, arising from the Yukawa couplings 
of RH neutrinos to LH leptons, and the Majorana mass matrix $M_N$ of 
the RH neutrinos, are present in the neutrino sector. 
Several possibilities can be considered
 that all can lead to the same results for lepton mixing:  
$m_D$ is invariant under the entire flavor (and CP) symmetry, 
while $M_N$ enjoys 
the residual symmetry $G_\nu$; or $M_N$ is invariant 
under the entire flavor (and CP) symmetry and $m_D$ only under $G_\nu$; or
both mass matrices, $m_D$ and $M_N$, are only constrained by $G_\nu$. 
In the scenarios we discuss here, we focus on the first 
\cite{Hagedorn:2009jy,Hagedorn:2016lva} 
and the third
possibility \cite{Mohapatra:2005ra,Mohapatra:2015gwa}.  
The second possibility can also be very interesting, if resonant leptogenesis    
is the mechanism generating $Y_{\Delta B}$ \cite{leptogenesis:A03}.

\subsection{Predictions for leptogenesis}

We first make some general observations regarding the size of the CP asymmetries $\epsilon_{k (\alpha)}$ and 
their phase dependence.  
Then, we discuss examples for flavored \cite{Mohapatra:2015gwa} as well as unflavored leptogenesis \cite{Mohapatra:2005ra,Hagedorn:2009jy,Hagedorn:2016lva} in scenarios with flavor symmetries  $G_f$   \cite{Mohapatra:2005ra,Hagedorn:2009jy}
 as well as in scenarios with flavor and CP symmetries \cite{Mohapatra:2015gwa,Hagedorn:2016lva} and under the assumption of different residual symmetries $G_e$ and $G_\nu$. 
 For concreteness, we consider non-supersymmetric theories. However, all results can be easily adapted to supersymmetric models.
 Moreover, we focus on scenarios in which RH neutrino masses are not degenerate and hence results may differ for resonant leptogenesis.

\subsubsection{General observations}

First, we discuss the case in which LH leptons and RH neutrinos form irreducible triplets ${\bf 3}$ of a flavor group $G_f$ (and there might or might not be a CP symmetry present), since this is the most common situation 
for non-abelian $G_f$.\footnote{For the reader interested in other cases we refer to \cite{Bertuzzo:2009im}.} The Dirac neutrino mass matrix $m_D$ can be expanded, like all other mass matrices, in terms of the (small, positive)
flavor (and CP) symmetry breaking parameter $\kappa$
\begin{equation}
m_D = m_D^0 + \delta m_D + \mathcal{O}(\kappa^2\,v)
\end{equation}
with $m_D^0\sim \mathcal {O}(v)$ and $\delta m_D \sim \mathcal{O}(\kappa\,v)$. Typically $\kappa$ is of order $10^{-3} \lesssim \kappa \lesssim 10^{-1}$.

The combination $\hat{m}_D^\dagger \hat{m}_D$ can be expanded as
\begin{equation}
\label{mDdagmDexp}
\hat{m}_D^\dagger \hat{m}_D \approx U_R^\dagger \, \left( (m_D^0)^\dagger m_D^0 +  (m_D^0)^\dagger \, \delta m_D + (\delta m_D)^\dagger \,  m_D^0  \right) \, U_R \;.
\end{equation}
In the limit $\kappa \rightarrow 0$ this combination is diagonal in flavor space, since the first term on the right-hand side in \eref{mDdagmDexp} is proportional to the identity matrix.
 Hence, the CP asymmetries vanish for unflavored as well as flavored leptogenesis. Once corrections $ \delta m_D \sim \mathcal{O}(\kappa\, v)$ are included, we find in general for unflavored leptogenesis  
\begin{equation}
\epsilon_k   \sim \sum_{j \neq k} \mathrm{Im} \left( \left(\hat{m}_D^\dagger \hat{m}_D\right)_{kj}^2 \right) \propto \kappa^2
\end{equation}
and for flavored leptogenesis
\begin{equation}
\label{flleptoeps}
\!\!\!\!\!\epsilon_{k \alpha} \sim \sum_{j \neq k} \mathrm{Im} \left( \left(\hat{m}_D^\dagger \hat{m}_D\right)_{\left[\tiny\begin{array}{c}kj\\jk\end{array}\right]\normalsize} (\hat{m}_D)_{\alpha k}^\ast (\hat{m}_D)_{\alpha j} \right) 
\propto \kappa \, .
\end{equation}
Realizations with specific flavor symmetries can be found in  \cite{Bertuzzo:2009im,Hagedorn:2016lva}.
 Note that the combination $\hat{m}_D^\dagger \hat{m}_D $ does not depend, up to order $\kappa$, on the phases present in $\delta m_D$, if the matrix $(m_D^0)^\dagger \, \delta m_D$
is symmetric (for further options see \cite{Hagedorn:2016lva}), since in this case the second and third term on the right-hand side in \eref{mDdagmDexp} are complex conjugated to each other.

If a $Z_2$ symmetry determines the relevant structure of $m_D$, either because it is a factor of the abelian flavor symmetry $G_f$
or $m_D^0$ vanishes and the leading term is given by $\delta m_D$, invariant under $Z_2$ only, the CP asymmetries $\epsilon_{k (\alpha)}$ do not vanish and are in general not suppressed by a 
small parameter \cite{Mohapatra:2005ra,Mohapatra:2015gwa}. 

If the theory possesses a CP symmetry that is left intact in the neutrino sector, it is true that the CP asymmetries $\epsilon_k$ vanish in the case of unflavored leptogenesis,
but $\epsilon_{k \alpha}$ can be non-zero and thus flavored leptogenesis can explain the baryon asymmetry $Y_{\Delta B}$. A proof of this statement can straightforwardly be obtained by observing that the two conditions
\begin{equation}
X ({\bf 3})^\ast \, m_D^\dagger m_D \, X ({\bf 3}) = (m_D^\dagger m_D)^\ast \;\;\; \mbox{and} \;\;\; X ({\bf 3}) \, M_N \, X ({\bf 3}) = M_N^\ast \; ,
\end{equation}
which express the invariance of $m_D$ and $M_N$ under the CP transformation $X ({\bf 3})$, require the following form of $m_D^\dagger m_D$ and $M_N$
\begin{eqnarray}
\label{CPonly}
\begin{split}
& m_D^\dagger m_D  \; =\; \Omega \, O_D \, \mathrm{diag} (m_{d \, i}^2) \, (\Omega \, O_D)^\dagger \,,\\
& M_N  \; =\; (\Omega \, O_R \, K_R)^\ast \, M_N^{\rm diag} \, (\Omega \, O_R \, K_R)^\dagger
\end{split}
\end{eqnarray}
where $\Omega$ is unitary and determined by $X ({\bf 3})$, 
$X ({\bf 3}) = \Omega \, \Omega^T$, $O_{D, R}$ are three-by-three 
rotation matrices, $m_{d \, i}^2 > 0$, $M_N^{\rm diag}$ contains the RH neutrino masses $M_k$ as diagonal entries and 
 $K_R$ is a diagonal matrix 
 with $\pm 1$ and $\pm i$, accounting for the CP parities of the RH neutrinos. 
Consequently, 
the application of $U_R = \Omega \, O_R \, K_R$ to $m_D^\dagger m_D$ shows 
that the off-diagonal elements of  $\hat{m}_D^\dagger \hat{m}_D$ can at most carry a factor $\pm i$ arising from $K_R$. Similar observations are also reported in \cite{Chen:2016ptr}. 

We exemplify these different situations in the following with well-known examples taken from the literature \cite{Hagedorn:2009jy,Mohapatra:2005ra,Hagedorn:2016lva,Mohapatra:2015gwa}.

\subsubsection{Leptogenesis in scenarios with flavor symmetries}

We present examples which show the predictive power of different types of flavor and residual symmetries
with regard to the CP asymmetries $\epsilon_k$. 

\paragraph{}

\noindent\textit{Case of residual symmetry $Z_2$ in neutrino sector.}
We consider an example with $\mu-\tau$ exchange symmetry  among neutrinos \cite{Fukuyama:1997ky,Mohapatra:1998ka} and charged leptons invariant under a discrete or continuous charged lepton flavor symmetry.
In \cite{Mohapatra:2004hta} the results for unflavored leptogenesis have been studied for a strongly hierarchical RH neutrino mass spectrum and with two or three RH neutrinos.
It has been shown that for two RH neutrinos the CP asymmetry $\epsilon_1$ vanishes in the limit of exact $\mu-\tau$ exchange symmetry, while for three RH neutrinos a non-zero
CP asymmetry is obtained. The latter result has also been found in \cite{Grimus:2003sf,Mohapatra:2005ra}. For $M_1 \ll M_{2,3}$ it is known that the 
CP asymmetry $\epsilon_1$, arising from the decay of the lightest RH neutrino $\mathcal{N}_1$, can be written as 
\begin{equation}
\epsilon_1 \approx - \frac{3 \, M_1}{8\pi v^2}\frac{\mathrm{Im} \Big(\left( \hat{m}_D^T 
M_\nu^\ast \hat{m}_D \right)_{11} \Big)}{(\hat{m}^\dagger_D \hat{m}_D)_{11}} = 
\frac{3 \, M_1}{8\pi v^2}\frac{\mathrm{Im} \left( R_{12}^2 \, \Delta m_{21}^2 + R_{13}^2 \, \Delta m_{31}^2 \right)}{\left( \sum_j \left| R_{1j}\right|^2  m_j \right)} 
\,.
\end{equation}
Here we have 
approximated the loop function $f(x)$ in \eref{loopfun} for strongly hierarchical RH neutrino masses. 
The $R$ matrix of the Casas-Ibarra
parametrization \cite{Casas:2001sr} is defined as
\begin{equation}
\label{Rmat}
R = i\, (M_N^{\rm diag})^{-1/2} \, m_D^T \, U_{\nu}^\ast \, (M_\nu^{\rm diag})^{-1/2} 
\end{equation}
in the charged lepton mass basis. 
If the neutrino sector is invariant under $\mu-\tau$ exchange
symmetry, i.e. both, $m_D$ and $M_N$, are invariant under the application of $P^{\mu\tau}$,
the $R$ matrix has block structure with $R_{13}=R_{23}=R_{31}=R_{32}=0$. Then,
 $\epsilon_1$ is proportional to the solar mass squared difference only \cite{Mohapatra:2005ra}
\begin{equation}
\epsilon_1 \approx \frac{3 \, M_1}{8\pi v^2}\frac{\mathrm{Im} \left( \sin^2 z  \right) \, \Delta m_{21}^2}{\left( \left| \cos z \right|^2 \, m_1 + \left| \sin z \right|^2 \, m_2 \right)} 
\, 
\end{equation}
with $z$ being the complex angle parametrizing the $R$ matrix. 
Since vanishing reactor mixing angle $\theta_{13}$ is experimentally excluded to high degree, $\mu-\tau$ exchange symmetry cannot be exact in the charged lepton mass basis.
An example in which $\mu-\tau$ exchange symmetry is broken in the RH neutrino sector is provided in \cite{Mohapatra:2005ra}. The element $R_{13}$ of the $R$
matrix then becomes proportional to $\theta_{13}$, $R_{13} \propto \theta_{13}$, and thus the CP asymmetry $\epsilon_1$ can be cast into the form
\begin{equation}
\epsilon_1 \propto \left( c_1 \, \Delta m^2_{21} + c_2 \, \Delta m^2_{31} \, \theta^2_{13} \right)\,,
\end{equation}
with $c_{1,2}$ depending on, e.g.,~the elements $R_{ij}$ with $i,j=1,2$ of the $R$ matrix.

For figures of $Y_{\Delta B}$ in different scenarios with $\mu-\tau$ exchange symmetry in the neutrino sector see \cite{Grimus:2003sf,Mohapatra:2005ra}.

\paragraph{}

\noindent\textit{Case of residual symmetry $Z_2 \times Z_2$ in neutrino sector.}
We discuss results for a model with $G_f=A_4$ that leads to TB lepton mixing \cite{Altarelli:2005yx}. This model is originally formulated in a supersymmetric context.
However, this does not have an impact on the relation of the CP asymmetries $\epsilon_k$ to the low energy CP phases. 
 The residual symmetry $G_e=Z_3$ constrains
charged leptons to have a diagonal mass matrix, while the Dirac neutrino mass matrix $m_D$ is invariant under the entire flavor group and 
 the RH neutrino mass matrix $M_N$ has $G_\nu=Z_2 \times Z_2$ as symmetry. The actual form of $m_D$ and $M_N$ at leading order is
 \begin{equation}
m_D = m_D^0 = \overline{m} \, \left(
\begin{array}{ccc}
1 & 0 & 0\\
0 & 0 & 1\\
0 & 1 & 0
\end{array}
\right)
\;\;\; \mbox{and} \;\;\;
M_N = \left(
\begin{array}{ccc}
A +2 \, B & -B & -B\\
-B & 2 \, B & A-B\\
-B & A-B & 2 \, B
\end{array}
\right) 
\end{equation}
with $A$ and $B$ being complex mass parameters, while $\overline{m} > 0$ can be achieved via a field redefinition.
The light neutrino mass matrix arising from the type I seesaw mechanism leads to TB lepton mixing. Light and heavy neutrino masses $m_i$ and $M_k$ are strongly correlated, since the form of $m_D$
is trivial in flavor space, i.e.
\begin{equation}
\label{miMirel}
m_i = \frac{\overline{m}^2}{M_i} \,.
\end{equation}
In addition, they only depend on two complex parameters. They, hence, fulfill the following light neutrino mass sum rule involving the Majorana phases $\alpha$ and $\beta$  
\begin{equation}
\frac{1}{m_1} -\frac{e^{2 \, i \, \beta}}{m_3} = \frac{2 \, e^{2 \, i \, \alpha}}{m_2} \;.
\end{equation}
  Consequently, the mass spectra of light and heavy neutrinos are non-hierarchical and can have both orderings. 
  In turn, all three RH neutrinos contribute to the generation of $Y_{\Delta B}$. 
  The off-diagonal elements of $\hat{m}_D^\dagger \hat{m}_D$ vanish at leading order  
  and hence also the CP asymmetries $\epsilon_{k (\alpha)}$.
 
 In explicit models, however, the form of $m_D$ and $M_N$ as well as $M_e$ 
 receives corrections at the subleading level which break the residual symmetries $G_\nu$ and $G_e$, respectively.
  In the particular $A_4$ model the dominant correction $\delta m_D$ turns out to be invariant under $G_e$ 
 \begin{equation}
 \delta m_D = \overline{m}\, \left(
 \begin{array}{ccc}
  2 \, y_1 & 0 & 0\\
 0 & 0 & -y_1 - y_2\\
 0 & -y_1+y_2 & 0
 \end{array}
 \right) \, \kappa \;\; \mbox{with} \;\; y_{1,2} \;\; \mbox{complex.}
 \end{equation}
  Including $\delta m_D$ when computing $\epsilon_k$ in the case of unflavored leptogenesis, we find \cite{Hagedorn:2009jy} 
\begin{equation}
\epsilon_1\propto \kappa^2 \, \left( 6 \, \mathrm{Re} \left( y_1 \right)^2 \, f \left( \frac{m_1}{m_2} \right) \, \sin 2 \, \alpha +  \mathrm{Re} \left( y_2 \right)^2 \, f \left( \frac{m_1}{m_3} \right) \, \sin 2 \, \beta \right) 
 \end{equation}
 and similar results for $\epsilon_2$ and $\epsilon_3$.
  This shows that the CP asymmetries crucially depend on the two Majorana phases $\alpha$ and $\beta$ and, since the matrix $(m_D^0)^\dagger \, \delta m_D$ is symmetric,
 the imaginary parts of the parameters $y_{1,2}$ of the correction do not enter at  $\mathcal{O}(\kappa^2)$. $f(x)$ is the loop function in \eref{loopfun} and its argument can be written in terms of light neutrino masses
 with the help of \eref{miMirel}.
The corrections to the mass matrices affect in general also the predictions for lepton mixing. However, since $\kappa$ is small, these corrections only have a minor impact \cite{Altarelli:2005yx}. 
 
In order to compute $Y_{\Delta B}$ the efficiency factors $\eta_{ij}$ have to be taken into account \cite{Hagedorn:2009jy}. In this type of model RH neutrinos couple at leading order
to orthogonal linear combinations of lepton flavors and thus the contributions $Y_{\Delta B, k}$ from the decay of the three different RH neutrinos $\mathcal{N}_k$ to $Y_{\Delta B}$ can be summed incoherently.
For figures of $Y_{\Delta B}$ see \cite{Hagedorn:2009jy}.

 \subsubsection{Leptogenesis in scenarios with flavor and CP symmetries}

As has been argued, imposing a CP symmetry in addition to a flavor symmetry increases the predictive power
of the approaches regarding  CP phases so that it can become possible to correlate the sign of $Y_{\Delta B}$
with the low energy CP phases. We exemplify this in the following with the help of two concrete scenarios taken from \cite{Mohapatra:2015gwa,Hagedorn:2016lva}.

\paragraph{}

\noindent{\it Case of residual CP symmetry in neutrino sector.} We discuss a setup \cite{Mohapatra:2015gwa} in which the charged lepton mass matrix is diagonal due to the choice of $G_e$ and $M_\nu$ is constrained
by $\mu-\tau$ reflection symmetry \cite{Harrison:2002kp}, $X ({\bf 3})= P^{\mu\tau}$. In particular, the Majorana mass matrix $M_N$ of the RH neutrinos
and the Dirac neutrino mass matrix $m_D$, both are invariant under $\mu-\tau$ reflection symmetry. As already shown above for a general CP symmetry,
preserved in the neutrino sector, the CP asymmetries $\epsilon_k$ vanish in the case of unflavored leptogenesis. This observation has also been made in 
 \cite{Grimus:2003yn}. In contrast, $\epsilon_{k \alpha}$
are normally non-zero. Thus, we consider in the following RH neutrinos with masses below $10^{12}$ GeV so that flavor effects become relevant when computing
the CP asymmetries. 

We note that from 
\begin{equation}
X ({\bf 3})^\ast \, m_D \, X ({\bf 3}) = P^{\mu\tau} \, m_D \, P^{\mu\tau}= m_D^\ast
\end{equation}
together with \eref{CPonly} the following relation can be derived for $\hat{m}_D$
\begin{equation}
\label{mDstarmD}
\hat{m}_D^\ast =  P^{\mu\tau}  \, \hat{m}_D \, K_R^2 \,,
\end{equation}
$K_R^2$ being a diagonal matrix with entries $\pm 1$.

Assuming that RH neutrinos are strongly hierarchical, $M_1 \ll M_{2,3}$, we only consider 
the CP asymmetries $\epsilon_{1 \alpha}$, generated in the decay of the RH neutrino $\mathcal{N}_1$. 
In order to evaluate the expression in \eref{epskagen} for $\epsilon_{1 \alpha}$ we use \eref{mDstarmD},
taking into account all possible choices for $K_R^2$. We find that 
\begin{equation}
\epsilon_{1 e} = 0 \, ,
\end{equation}
whereas 
\begin{equation}
\label{epsmutau}
\epsilon_{1 \mu} = - \epsilon_{1 \tau} \neq 0
\end{equation}
in general. This leads to vanishing $\epsilon_1$, as expected.
Since $\epsilon_{1 \mu}$ and $\epsilon_{1 \tau}$ have opposite
sign, but equal magnitude, $Y_{\Delta B}$ can only be explained, if we consider the regime in which the $\tau$ flavor alone 
can be distinguished by its fast Yukawa interaction in the primordial plasma. This leads to $10^9$ GeV as lower bound  
on the RH neutrino masses.

The source of CP violation that leads to non-vanishing CP asymmetries $\epsilon_{1 \mu}$ and $\epsilon_{1 \tau}$ can be
traced back to the maximal Dirac phase $\delta$. In order to see this consider
\begin{eqnarray}
\begin{split}
\epsilon_{1 \alpha}& \approx\; - \frac{3 \, M_1}{8\pi v^2}\frac{\mathrm{Im} \Big( \left(\hat{m}_D^T 
M_\nu^\ast \right)_{1 \alpha} (\hat{m}_D)_{\alpha 1} \Big)}{(\hat{m}^\dagger_D \hat{m}_D)_{11}}
\\ \label{epsRmatCP}
&=\; \frac{3 \, M_1}{8\pi v^2}\frac{\sum_{i,j} m_i \, \sqrt{m_i \, m_j} \,  \mathrm{Im} \left( R_{1 i} \, R_{1 j} \, U^\ast_{\nu, \alpha i} \, U_{\nu, \alpha j} \right)}{\left( \sum_k \left| R_{1k}\right|^2 \, m_k \right)} \,.
\end{split}
\end{eqnarray}
Furthermore, we note that the $R$ matrix fulfills 
\begin{equation}
\label{Rcon}
R^\ast = - K_R^2 \, R \, K_\nu^2\,,
\end{equation}
with $K_\nu$ encoding the CP parities of the light neutrinos.
Thus, $K_\nu^2$ is a diagonal matrix with entries $\pm 1$. 
This equation can be derived using \eref{mDstarmD} together 
with the constrained form of $U_{\nu}$. 
 Given \eref{Rcon}
the elements of the $R$ matrix can only be real or imaginary, 
i.e.~$R$ can be written as
\begin{equation}
 R = i\, K_R^\ast \, R^{(0)} \, K_\nu\,,
 \end{equation}
with $R^{(0)}$ being a real matrix. Plugging this information in \eref{epsRmatCP}, one can see that $\epsilon_{1 \tau}$ (and hence also $\epsilon_{1 \mu}$) 
is proportional to the Jarlskog invariant $J_{\rm CP}$ \cite{Jarlskog:1985ht}  and is consequently sourced from the low energy CP phase $\delta$, that is maximal.

\begin{figure}[t!]
\parbox{2.6in}{\includegraphics[width=2.5in]{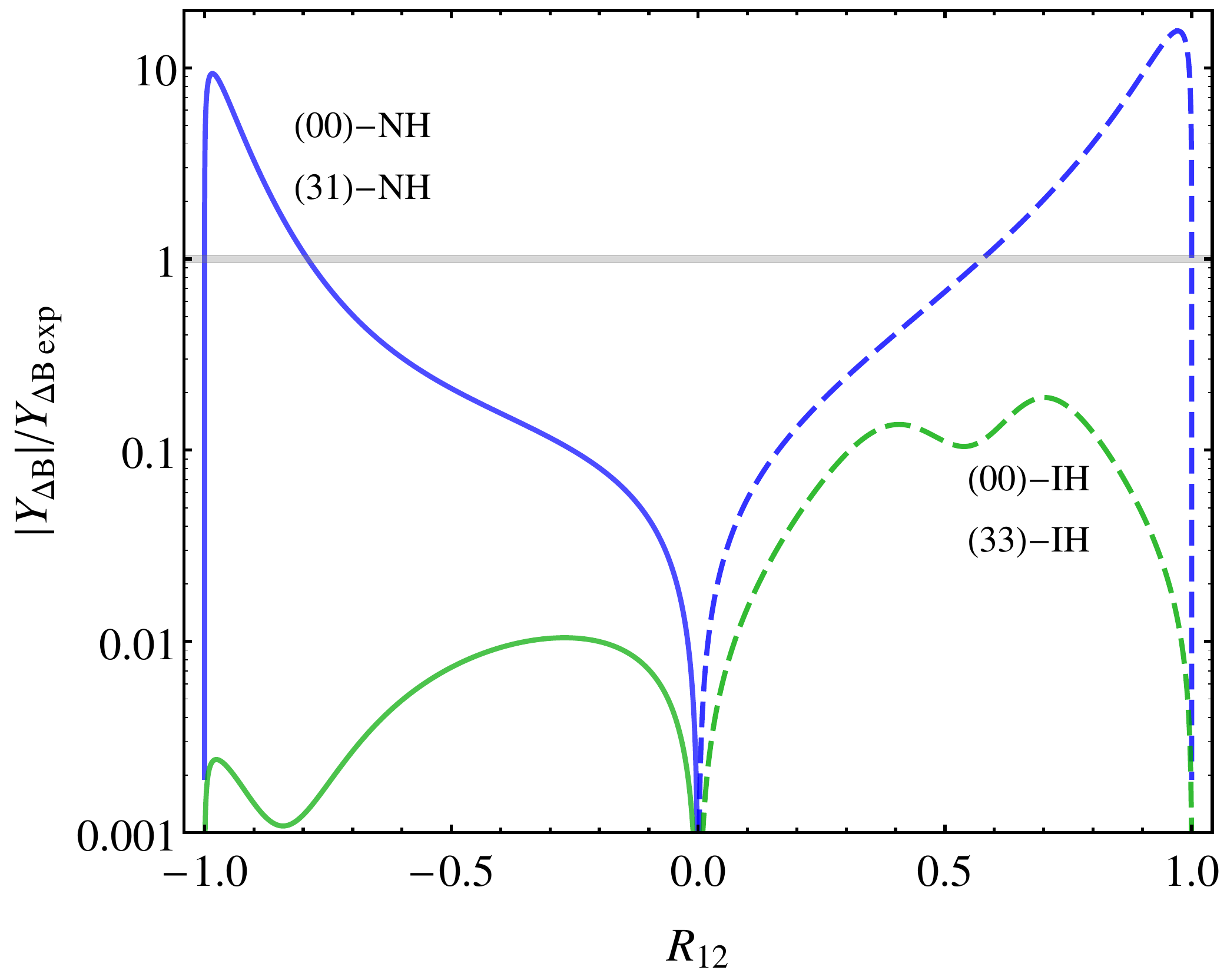}} \hspace{0.1in} \parbox{2.2in}{ 
 \caption{Prediction of magnitude of $Y_{\Delta B}$, normalized to experimental observation (indicated by the horizontal gray lines), versus the element $R_{12}$ of the $R$ matrix
in a scenario with $\mu-\tau$ reflection symmetry in the neutrino sector and a diagonal charged lepton mass matrix. Light neutrinos
follow NH and IH for blue and green curves. Solid (dashed) lines refer to cases with the Dirac
phase $\delta=3 \, \pi/2$ ($\delta=\pi/2$) and $Y_{\Delta B} > 0$. $(ij)$ indicate different choices of $K_R$ and $K_\nu$. Figure adapted from \cite{Mohapatra:2015gwa}.\label{fig2YBvsR12}
}}
\end{figure}

In \fref{fig2YBvsR12} results for the magnitude of $Y_{\Delta B}$, normalized to the experimentally observed value, versus 
the element $R_{12}$ of the $R$ matrix are shown for neutrinos with NH and IH. We assume
that $M_3 \gg M_{1,2}$ so that the RH neutrino $\mathcal{N}_3$ decouples and in turn the $R$ matrix has block structure.
The RH neutrino mass $M_1$ is set to its maximal admitted value $M_1=10^{12}$ GeV. For a smaller $M_1$ mass the magnitude of $Y_{\Delta B}$
has to be appropriately rescaled. Results for the two different values of $\delta$, $\delta=\pi/2$ and $\delta=3 \, \pi/2$, and $Y_{\Delta B}$ positive 
are displayed in different line style. Different choices of $K_R$ and $K_\nu$ are considered for the different curves and are indicated with $(ij)$. 
As one can see, not all these choices lead to a large enough value of the magnitude of $Y_{\Delta B}$.
For further details and results for other combinations $(ij)$ see \cite{Mohapatra:2015gwa}.

\paragraph{}

\noindent{\it Case of residual symmetry $Z_2 \times CP$ in neutrino sector.} We present a scenario \cite{Hagedorn:2016lva} in which a flavor and a CP symmetry determine lepton mixing and the sign of $Y_{\Delta B}$. We choose in the following as $G_f$ one of the groups
$\Delta (3 \, n^2)$ and $\Delta (6 \, n^2)$. For details about these groups see \cite{Luhn:2007uq,Escobar:2008vc}.
 The CP symmetry depends in general on one or two parameters.  
The residual symmetries are $G_e=Z_3$ and $G_\nu=Z_2 \times CP$. The basis of $G_f$ is chosen in such a way that $M_e$ is diagonal.
  Like in the discussed $A_4$ model \cite{Altarelli:2005yx}, the Dirac neutrino mass matrix $m_D$ respects all symmetries imposed on the theory, while the Majorana mass matrix $M_N$ of the RH neutrinos
  is only invariant under $G_\nu$
 \begin{equation}
m_D =  m_D^0 = \overline{m} \, \left(
\begin{array}{ccc}
1 & 0 & 0\\
0 & 1 & 0\\
0 & 0 & 1
\end{array}
\right) 
\quad \mbox{and} \quad
U_R^T \, M_N \, U_R = M_N^{\rm diag}\,,
\end{equation}
with $U_R = \Omega \, R_{ij} (\theta) \, K_R$ and $\overline{m}$ positive. The form of $\Omega$ is determined by $G_\nu$.
  The rotation $R_{ij} (\theta)$ in the $(ij)$-plane depends on $\theta$, $0 \leq \theta < \pi$, that is the only free parameter in $U_R$. The RH neutrino masses $M_k$ are also unconstrained. 
Applying the type I seesaw mechanism, we see that the lepton mixing matrix is given by $U_R$ and, thus, $\theta$ is fixed by the constraint to accommodate 
the measured lepton mixing angles. The light neutrino masses $m_i$ are, like in the $A_4$ model, inversely proportional to $M_i$, see \eref{miMirel}.
As expected, the CP asymmetries $\epsilon_{k (\alpha)}$ vanish at this level, because $(\hat{m}_D^\dagger \hat{m}_D)_{ij}=0$ for $i\neq j$, 
and corrections are needed for $\epsilon_{k (\alpha)} \neq 0$. Similar to the $A_4$ model,
we consider the dominant correction $\delta m_D$ to be invariant under $G_e$, the residual symmetry in the charged lepton sector. The form of $\delta m_D$ is then
\begin{equation}
\delta m_D = \overline{m} \, \left( \begin{array}{ccc}
 \frac{2}{\sqrt{3}} \, z_1 & 0 & 0\\
 0 & - \frac{1}{\sqrt{3}} \, z_1 - z_2 & 0\\
 0 & 0 & - \frac{1}{\sqrt{3}} \, z_1 + z_2
\end{array}
\right)  \kappa
\quad \mbox{with} \quad
 z_{1,2} \quad \mbox{complex.}
\end{equation}
It is convenient to parametrize the real parts of $z_{1,2}$ as $\mathrm{Re} (z_1) = z\, \cos\zeta$ and $\mathrm{Re} (z_2) = z\, \sin\zeta$, with $z > 0$ and $0 \leq \zeta < 2 \, \pi$.
In \cite{Hagedorn:2014wha,Ding:2014ora} it has been shown that four different types of mixing patterns, i.e.~$U_R$, exist and we focus in the following on Case 1) and Case 3 b.1) of \cite{Hagedorn:2014wha}. 

\begin{table}[t!]
\tbl{Prediction of CP phases $\alpha$, $\beta^\prime$ and $\delta$
 from $\Delta (6 \cdot 8^2)$ and $CP (s)$, 
  $G_e=Z_3$, $G_\nu=Z_2 (4) \times CP (s)$. $K_R$ is trivial \cite{Hagedorn:2014wha,Hagedorn:2016lva}. 
  }
{\begin{tabular}{l|ccccc} \toprule
$s \;\;$ & &&$\; \sin 2 \, \alpha=\sin 2\, \beta^\prime$&&$\sin\delta$\\
\hline
$1 \;\;\;\;$& &&$-1/\sqrt{2}$ && $\pm 0.936$\\
$2 \;\;\;\;$& &&$1$ && $-0.739$\\
$4 \;\;\;\;$& &&$0$ && $\mp 1$\\
\hline
\end{tabular}}\label{tab1:CPphases}
 \end{table}

Case 1) predicts only the Majorana phase
$\alpha$ to be non-trivial
\begin{equation}
| \sin 2\, \alpha | = \bigg|\sin \left( \frac{6 \, \pi \, s}{n} \right) \bigg| \,,\quad \sin 2 \, \beta^\prime=0 \,,\quad \; \sin \delta =0 \,.
\end{equation}
The integer parameter $s$ characterizes the CP symmetry $CP (s)$ and lies in the interval $0 \leq s \leq n-1$. The CP asymmetries $\epsilon_k$ have a very simple form and we find \cite{Hagedorn:2016lva}
\begin{equation}
\epsilon_1 \propto \kappa^2 \, \cos^2 \left( \theta+\zeta \right) \, f \, \left( \frac{m_1}{m_2} \right) \, \sin \left( \frac{6 \, \pi \, s}{n} \right) 
\propto \kappa^2 \, f \, \left( \frac{m_1}{m_2} \right) \,  I_1 \left( \theta \rightarrow \theta + \zeta \right) 
\end{equation}
and similar expressions for $\epsilon_2$ and $\epsilon_3$.
This formula shows that the sign of $\epsilon_1$ only depends on $s$, and hence $\alpha$, and the light neutrino mass spectrum through the sign of the loop function $f(x)$. 
 Furthermore, we recognize that the terms in $\epsilon_k$ have the same structure as the Majorana CP invariants $I_i$, $i=1, 2, 3$ (for the definition of $I_i$ see \cite{Branco:1986gr,Nieves:1987pp}). As shown in \cite{Hagedorn:2016lva}, this not only
 holds for lepton mixing patterns derived from $G_f$ and CP, but can also be found if a generic form of the lepton mixing matrix is considered.

Case 3 b.1) has a much richer phenomenology, since all CP phases take in general CP-violating values. We focus on the flavor group $\Delta (6 \cdot 8^2)$ with the index $n=8$.
The residual symmetry $G_\nu$ is characterized by two integer parameters $m$ and $s$, $0 \leq m, \, s \leq n-1$. The symmetry $Z_2 (m)$  is fixed by choosing $m=n/2=4$
such that the solar mixing angle is accommodated well. As a consequence, the sines
of the Majorana phases coincide in size and also in sign, if $K_R$ is trivial, and they depend only on the chosen CP symmetry $CP (s)$ 
\begin{equation}
 | \sin 2\, \alpha | =  | \sin 2\, \beta^\prime | = \bigg|\sin \left( \frac{6 \, \pi \, s}{n} \right) \bigg| \,.
 \end{equation}
Requiring that all three lepton mixing angles agree with the experimental data at the $3\, \sigma$ level or better, 
  a lower bound on the size of the sine of the Dirac phase $\delta$ can be derived 
  \begin{equation}  
  | \sin \delta | \gtrsim 0.71 \,.
 \end{equation}
 For the admitted values of $s$, $s=1$, $s=2$ and $s=4$, the CP phases are predicted as in \tref{tab1:CPphases}.
 Since for $s=1$ and $s=4$ both, $\sin^2 \theta_{23}$ and $1-\sin^2 \theta_{23}$, lie in the experimentally preferred $3 \, \sigma$ range for $\sin^2 \theta_{23}$,
 $\sin\delta$ can be either positive or negative.

\begin{figure}[t!]
\parbox{2.8in}{\includegraphics[width=2.7in]{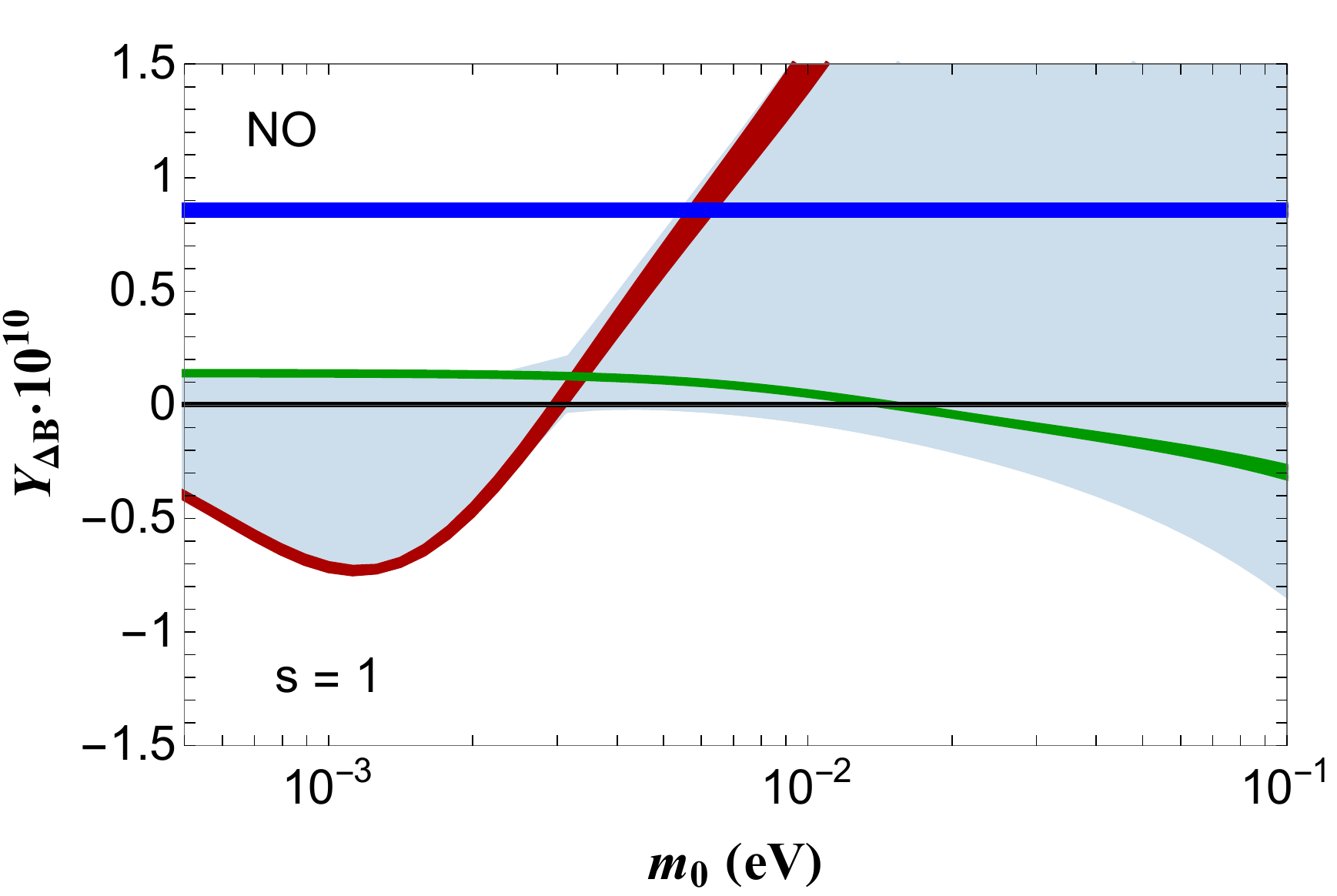}} \hspace{0.01in} \parbox{2in}{  
\caption{Prediction of $Y_{\Delta B}$ \mathversion{normal} versus  the lightest neutrino mass $m_0$ from $\Delta (6 \cdot 8^2)$ and $CP (1)$, 
  $G_e=Z_3$, $G_\nu=Z_2 (4) \times CP (1)$. $K_R$ is trivial. Light neutrino masses follow NO. Different colors refer to different $\zeta$. 
The horizontal blue band indicates the experimentally preferred $3 \, \sigma$ range of $Y_{\Delta B}$. Figure adapted from \cite{Hagedorn:2016lva}.\label{fig3YBvsm0}
}}
\end{figure}

   The prediction of $Y_{\Delta B}$ versus the lightest neutrino mass $m_0$ for the choice $s=1$  is shown in \fref{fig3YBvsm0}. One sees that for small $m_0$, $m_0 \lesssim 3 \times 10^{-3} \, \mathrm{eV}$, 
negative $Y_{\Delta B}$ is preferred, while larger values of $m_0$ strongly favor positive $Y_{\Delta B}$. For the choice $s=2$, predicting a different sign for $\sin 2 \, \alpha$ and $\sin 2 \, \beta^\prime$ than  $s=1$, see \tref{tab1:CPphases},  
the sign of $Y_{\Delta B}$ changes accordingly, i.e.~for $CP(2)$ positive $Y_{\Delta B}$ is achieved for small $m_0$, $m_0 \lesssim 4 \times 10^{-3} \, \mathrm{eV}$.
 For the choice $CP (4)$ in which both Majorana phases are CP-conserving, while the Dirac phase $\delta$
is maximal,\footnote{These results for the CP phases are also obtained in scenarios with $\mu-\tau$ reflection symmetry, see \sref{sec4.1}.} the sign of $Y_{\Delta B}$ cannot be predicted, since it is proportional to $\sin 2 \, \zeta$ at leading order in $\kappa$, i.e.~it depends on the relative sign of the real parts of the parameters in $\delta m_D$. 

Flavored leptogenesis can also be considered in a scenario in which residual flavor and CP symmetries constrain the neutrino sector, as mentioned in \cite{Hagedorn:2016lva} and studied in more detail in \cite{Chen:2016ptr}.

\mathversion{bold}
\subsection{Predictions for neutrinoless double beta decay}
\mathversion{normal}

One combination of the Majorana phases can be tested in $\betabeta$ decay. 
 The  effective Majorana mass $\left|m_{\beta\beta}\right|$ depends on the CP phases $\alpha$ and $\beta^{(\prime)}$, the light neutrino masses, and the solar and reactor
mixing angle, see \eref{effmass2}.

In the example of $\mu-\tau$ reflection symmetry in the neutrino sector and a diagonal charged lepton mass matrix both Majorana phases are CP-conserving. In such a situation 
 both, minimal and maximal, attainable values of $\left|m_{\beta\beta}\right|$ for a given lightest neutrino mass $m_0$ can be obtained, independent of the neutrino mass ordering. 
 These results are shown in \fref{Fig1} (darker bands) and can also be found in \cite{Mohapatra:2015gwa}.
 Note that $\betabeta$  decay experiments cannot distinguish
between the CP-conserving scenario and the predictions of   $\mu-\tau$  reflection symmetry,
and information on the Dirac phase $\delta$ is needed, since in
the first case
$\delta$ is CP-conserving, while in the second one it is maximal.

\begin{figure}[t!]
\parbox{1.9in}{\includegraphics[width=2.65in]{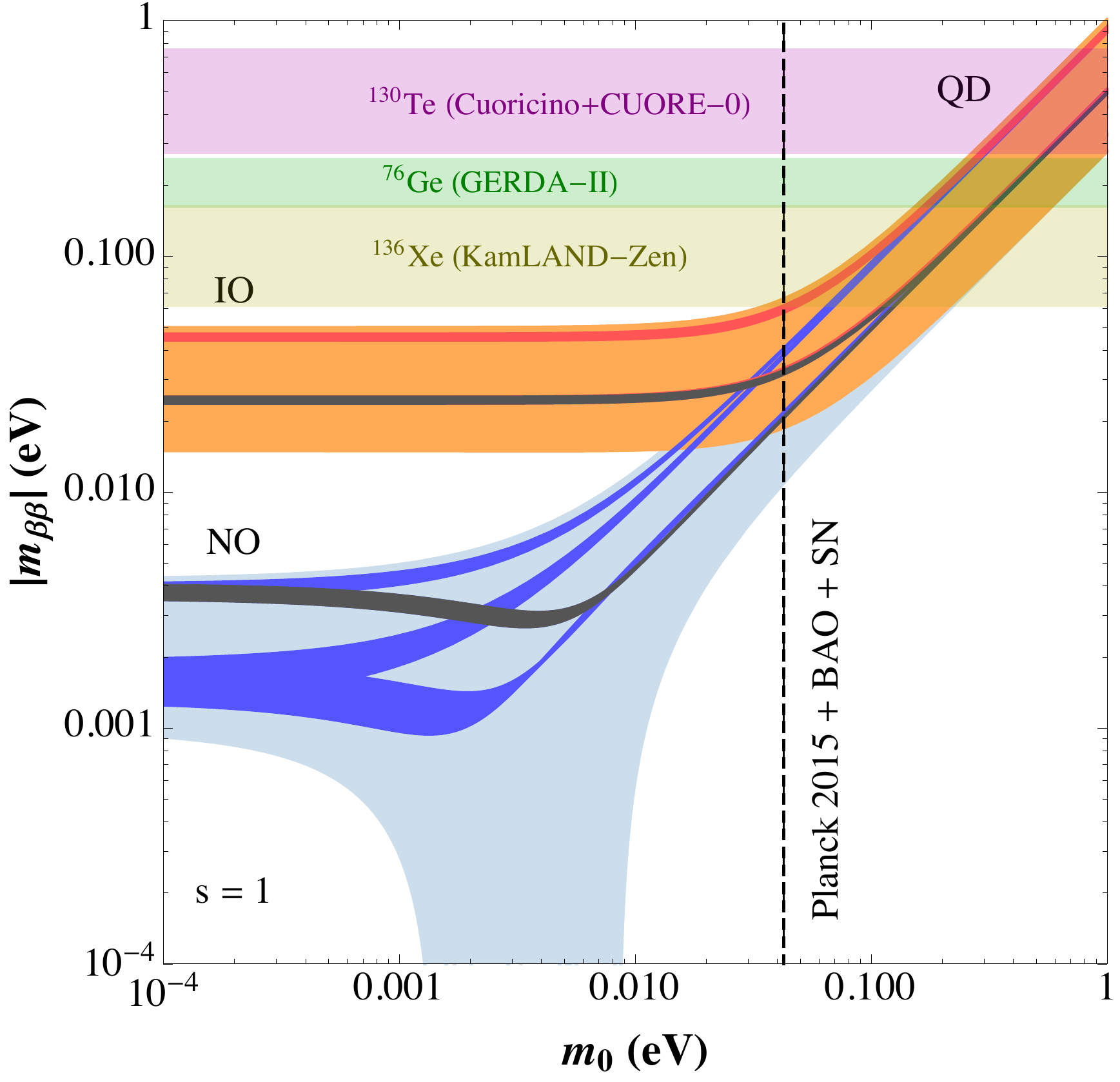}} \hspace{1.0in}\parbox{2.0in}{ \caption{Prediction
of $\left|m_{\beta\beta}\right|$ \mathversion{normal} versus $m_0$ from $\Delta (6 \cdot 8^2)$ and $CP (1)$, 
  $G_e=Z_3$, $G_\nu=Z_2 (4) \times CP (1)$. The orange (blue) areas
refer to IO (NO) light neutrino masses. 
This approach strongly restricts the allowed values of $\left|m_{\beta\beta}\right|$ (darker colors) compared to the experimental constraints on the lepton mixing parameters alone (lighter colors).
The choice $K_R$ trivial is highlighted in dark grey. Exclusion limits from searches for $\betabeta$ decay and from cosmology are also shown. Figure adapted from \cite{Hagedorn:2016lva}. \label{fig4meevsm0}
}}
\end{figure}

The approach with a residual flavor and a CP symmetry in the neutrino sector is particularly powerful in constraining $\left|m_{\beta\beta}\right|$, since all CP phases and lepton mixing angles 
are highly correlated and predicted to lie in a small range, once the free parameter $\theta$ is fixed. We use the same flavor and CP symmetry as in the example for unflavored leptogenesis, see \fref{fig3YBvsm0}. 
 The numerical values of the CP phases are shown in Table~\ref{tab1:CPphases} and in \fref{fig4meevsm0} we display the prediction of $\left|m_{\beta\beta}\right|$ versus $m_0$ for neutrino masses with NO and IO for the choice $s=1$\cite{Hagedorn:2016lva}.
The predictive power of this approach is underlined by comparing the very limited ranges of $\left|m_{\beta\beta}\right|$ in this case (darker colors in \fref{fig4meevsm0})
with the ranges of $\left|m_{\beta\beta}\right|$ allowed by the experimental constraints on the lepton mixing parameters alone (lighter colors in \fref{fig4meevsm0}). In particular,
we see that for NO $\left|m_{\beta\beta}\right|$ has a non-trivial lower bound and for IO the minimum attainable value of $\left|m_{\beta\beta}\right|$ is larger than $0.023 \, \mathrm{eV}$. For certain choices of $K_R$
 we find, indeed, that $\left|m_{\beta\beta}\right|$ is close to its upper limit. 
 
 Other examples of predictions of the Dirac and Majorana phases 
as well as of $\left|m_{\beta\beta}\right|$ in theories with combined 
flavor and CP symmetries can be found, e.g., in 
\cite{King:2014rwa, Li:2015jxa}.

\section{Summary and conclusions}

In this chapter we have reviewed the experimental status of neutrino data and future perspectives of measuring CP violation in the lepton sector. We have pointed out how a possible discovery of CP-violating effects in neutrino oscillations and/or the observation of $\betabeta$ decay may reveal a deep connection between the origin of neutrino masses and lepton mixing and the production of the cosmological baryon asymmetry $Y_{\Delta B}$ via the leptogenesis  mechanism.

We have presented different approaches in order to explain the observed lepton mixing angles with the help of flavor (and CP) symmetries
and have shown which predictions for the yet-to-be-measured CP phases $\delta$, $\alpha$, $\beta^{(\prime)}$ can be made. We have then discussed
how such approaches can be implemented in scenarios in which light neutrino masses arise from the type I seesaw mechanism. In
these scenarios  $Y_{\Delta B}$ can be generated via the decay of heavy RH neutrinos. We have shown results for different
 variants and have elucidated in which ones and how the sign of $Y_{\Delta B}$ can be related to predictions for the CP phases $\delta$, $\alpha$, $\beta^{(\prime)}$.
 
We have focussed on the type I seesaw mechanism in this chapter. It is, however, also possible to implement the approaches with
flavor (and CP) symmetries in scenarios with other generation mechanisms for neutrino masses. In case such scenarios 
 offer a way to generate $Y_{\Delta B}$, flavor (and CP) symmetries can leave an imprint on the results for $Y_{\Delta B}$ as well. 
  An analysis of leptogenesis in scenarios with type II \cite{Magg:1980ut,Schechter:1980gr,Mohapatra:1980yp} and type III \cite{Foot:1988aq} seesaw mechanisms 
and the flavor symmetry $S_4$ can be found in \cite{Bazzocchi:2009da}.

\section*{Acknowledgments}
C.H. would like to thank P.S.~Bhupal Dev and Jacobo Lopez-Pavon for discussions
on related issues.
The work of C.H. and E.M. is supported by the Danish National Research Foundation Grant, DNRF-90. 
The work of R.N.M. was supported by the US National Science Foundation under Grant No. PHY1620074.
C.C.N acknowledges support by Brazilian Fapesp grant 2014/19164-6 and 2013/22079-8,
and CNPq 308578/2016-3.
The work  of S.T.P. was supported in part by the INFN
program on Theoretical Astroparticle Physics (TASP),
by the European Union Horizon 2020 research and innovation program under the  Marie Sk\l{}odowska-Curie grants 674896 and 690575, and by the World Premier International Research Center Initiative (WPI Initiative), MEXT, Japan.
This work has been initiated at the Munich Institute for Astro- and Particle Physics (MIAPP) of the DFG cluster of excellence ``Origin and Structure of the Universe''.

\bibliographystyle{ws-rv-van-mod2}

\bibliography{chapter-06}

\end{document}